\documentclass[10pt,preprint2]{aastex}
\usepackage{epsfig}
\usepackage{natbib}

\newcommand{\oiii}{[OIII]}
\newcommand{\vvmax}{$V/V_{\rm max}$}
\newcommand{\invvmax}{$1/V_{\rm max}$}

\newcommand{\rmoiii}{{\rm [OIII]}}
\newcommand{\vmax}{$V_{\rm max}$}
\newcommand{\loiii}{$L_\rmoiii$}
\newcommand{\beq}{\begin{equation}}
\newcommand{\eeq}{\end{equation}}
\newcommand{\oiiil}{[OIII]5007}
\newcommand{\feii}{FeII}
\newcommand{\hb}{H$\beta$}

\newcommand{\mvvmax}{$\langle V/V_{\rm max}\rangle$}

\newcommand{\za}{$z\le 0.30$}
\newcommand{\zb}{$0.30<z\le 0.50$}
\newcommand{\zc}{$0.50<z<0.83$}
\newcommand{\Mi}{M_{2500}}
\newcommand{\pzl}{p_{\rm RD}(z,L_{\rm [OIII]})}
\newcommand{\prad}{p_{\rm RD}}
\newcommand{\hahb}{H$\alpha/$H$\beta$}
\newcommand{\hbhg}{H$\beta$/H$\gamma$}

\begin{document}
\shorttitle{Space Density of Type 2 Quasars}
\title{Space Density of Optically-Selected Type 2 Quasars}
\author{
Reinabelle Reyes\altaffilmark{1}, 
Nadia L. Zakamska\altaffilmark{2,3}, 
Michael A. Strauss\altaffilmark{1}, 
Joshua Green\altaffilmark{1}, 
Julian H. Krolik\altaffilmark{4}, 
Yue Shen\altaffilmark{1},
Gordon T. Richards\altaffilmark{5},
Scott F. Anderson\altaffilmark{6},
Donald P. Schneider\altaffilmark{7}\\
\altaffiltext{1}{Princeton University Observatory, Peyton Hall, Princeton, New Jersey 08544.}
\altaffiltext{2}{Institute for Advanced Study, Einstein Drive, Princeton, NJ 08540.}
\altaffiltext{3}{Spitzer Fellow, John N. Bahcall Fellow.}
\altaffiltext{4}{Department of Physics and Astronomy, Johns Hopkins University, 3400 North Charles Street, Baltimore, MD 21218-2686.}
\altaffiltext{5}{Department of Physics, Drexel University, 3141 Chestnut Street, Philadelphia, PA 19104.}
\altaffiltext{6}{Astronomy Department, Box 351580, University of Washington, Seattle, WA 98195}
\altaffiltext{7}{Department of Astronomy and Astrophysics, Pennsylvania State University, 525 Davey Laboratory, University Park, PA 16802, USA.}
} 

\begin{abstract}
Type 2 quasars are luminous active galactic nuclei (AGN) whose central regions are obscured by large amounts of gas and dust. In this paper, we present a catalog of type 2 quasars from the Sloan Digital Sky Survey (SDSS), selected based on their optical emission lines. The catalog contains 887 objects with redshifts $z<0.83$; this is six times larger than the previous version and is by far the largest sample of type 2 quasars in the literature. We derive the \oiiil {} luminosity function for $10^{8.3} L_{\odot} < L_{\rm [OIII]} < 10^{10} L_{\odot}$ (corresponding to intrinsic luminosities up to $M[2500{\rm \AA}]\simeq -28$ mag or bolometric luminosities up to $4\times 10^{47}$ erg s$^{-1}$). This luminosity function provides robust lower limits to the actual space density of obscured quasars, due to our selection criteria, the details of the spectroscopic target selection, as well as other effects. We derive the equivalent luminosity function for the complete sample of type 1 (unobscured) quasars and determine the ratio of type 2 to type 1 quasar number densities. Our data constrain this ratio to be at least $\sim 1.5:1$ for $10^{8.3} L_\odot <  L_{\rm [OIII]} < 10^{9.5} L_\odot$ at $z<0.3$, and at least $\sim 1.2:1$ for $L_{\rm [OIII]} \sim 10^{10} L_\odot$ at $0.3<z<0.83$. Type 2 quasars are at least as abundant as type 1 quasars in the relatively nearby Universe ($z\la 0.8$) for the highest luminosities.  
\end{abstract}

\section{Introduction}
\label{sec:intr}

Active galactic nuclei (AGN) can be classified as type 1 (unobscured) or type 2 (obscured) based on the presence or absence of broad hydrogen and helium emission lines in their optical spectra. Unification models of AGN attribute the distinction between these two types to a difference in the observer's viewing angle to a nucleus surrounded by non-isotropic obscuring material \citep{anto93, urry95}. These models are well-established for low-luminosity, nearby AGN. However, their applicability to high-luminosity AGN (i.e., quasars, classically defined to be sources with bolometric luminosities greater than $10^{45}$ erg s$^{-1}$) has long been controversial. Moreover, while it has long been known that obscured AGN dominate the low-luminosity population in the local Universe \citep{oste88, salz89, huch92, hao05b, simp05}, the situation is less clear for high-luminosity AGN. This is in part because these objects are much rarer and more difficult to sample, since the AGN luminosity function decreases steeply with luminosity. 

In this paper, we present a catalog of 887 type 2 quasars from the Sloan Digital Sky Survey (SDSS; \citealt{york00}), the largest sample of type 2 quasars in the literature to date. They are selected based on their optical emission lines, have redshifts $z< 0.83$ and \oiiil {} emission line luminosities extending to $\mbox{\loiii} \ga 10^{10} L_\odot$ (corresponding to intrinsic UV luminosity $M_{2500}\la -28$ mag). Multi-wavelength observations of the most luminous objects from a previous version of the sample (\citealt{zaka03}, hereafter Paper I) have confirmed that they have infrared luminosities up to and above $10^{46}$ erg s$^{-1}$, have spectral energy distributions expected of type 2 quasars \citep{zaka04, vign04, ptak06, vign06, zaka08}, and contain type 1 quasars in their centers revealed by polarimetric measurements \citep{zaka05, zaka06}. 

With this updated catalog drawn from roughly three times as much SDSS data, we can now sample the high-luminosity AGN population in sufficiently large numbers to draw quantitative conclusions. First, we derive the \oiiil {} luminosity function (LF) of type 2 quasars. Then, by directly comparing the space densities of type 2 and type 1 sources, we place robust lower limits on the fraction of obscured quasars in the local Universe as a function of \oiiil {} luminosity. Studying the space densities of different types of AGN provides strong constraints on the simplest AGN unification scenario as well as its modifications. Moreover, quantifying the obscured quasar population is essential for many applications, such as relating the present mass density of local black holes to the accretion history of the entire AGN population (e.g., \citealt{solt82,yu02,marc04}), understanding the origin of the cosmic X-ray background (e.g., \citealt{coma95,gill07}), and studying the effects of luminosity on AGN structure (e.g., \citealt{lawr91,urry95,hopk06}).

Our results are complementary to those derived from obscured quasars selected from hard X-ray surveys \citep{ueda03, szok04, barg05, mark05, trei06, beck06, sazo07} and mid-infrared color selection \citep{lacy05, ster05, mart06, poll07}.  For example, up to 20\% of hard X-ray selected AGN do not show any emission lines in their optical spectra \citep{rigb06}, and therefore would not be included in our sample. On the other hand, our sample would include Compton-thick objects that are missed in X-ray surveys, as long as their optical spectra meet our selection criteria.

The paper is structured as follows. In \S \ref{sec:samp}, we discuss our selection of type 2 quasars based on their optical emission lines and in \S\ref{sec:type2_lf}, we determine the \oiiil {} luminosity function from this sample. In \S\ref{sec:type1_lf}, we determine the equivalent luminosity function from a complete sample of type 1 quasars and in \S\ref{sec:type12}, we determine the ratio of type 2 to type 1 quasars using the derived luminosity functions. We discuss caveats and implications of these results in \S\ref{sec:discussion} and we conclude in \S\ref{sec:conclusions}. We adopt a `concordance' cosmology, $h = 0.7$, $\Omega_{\rm M}=0.3$, and $\Omega_\Lambda = 0.7$. We identify emission lines using air wavelengths, identify objects with J2000 coordinates, and use asinh magnitudes \citep{lupt99} corrected for Galactic extinction \citep{schl98}. We often express luminosities in units of solar luminosities $L_\odot = 3.826 \times 10^{33}$ erg s$^{-1}$.

\section{Type 2 Quasar Sample}
\label{sec:samp}

Our sample of type 2 quasars is selected from the SDSS spectroscopic database as objects with high-ionization, narrow emission lines, following Paper I. We describe SDSS data processing in $\S$\ref{subsec:sdss_data}, the targeting of objects for spectroscopy in SDSS in $\S$\ref{subsec:sdss_spec}, our spectroscopic selection criteria in $\S$\ref{subsec:sdss_selec} and measurement of the \oiiil {} emission line in \S\ref{subsec:type2_o3}.

\subsection{SDSS Data}
\label{subsec:sdss_data}

The SDSS has imaged $\sim\!$ 10,000 deg$^2$ of the sky with good astrometric and photometric calibration \citep{pier03, tuck06, smit02, ivez04, padm07} in the SDSS $ugriz$ filters \citep{fuku96, stou02} using a drift-scanning wide-field camera \citep{gunn98} on a dedicated 2.5 m telescope \citep{gunn06}. For each object the photometric pipeline returns various measures of flux in each band, such as Petrosian (1976), point spread function (PSF) and model magnitudes \citep{stou02}. 

A subset of objects from the imaging survey are targeted for spectroscopy and assigned to a series of plates containing 640 fibers each \citep{blan03}, each of which subtends 3$\arcsec$ on the sky; the spectra cover $3800-9200\mbox{\AA}$ with resolution of $1800<R<2100$ and sampling of $\simeq 2.4$ pixels per resolution element. The relative and absolute spectrophotometric calibration are good to $\sim 5\%$. Spectral flux errors per pixel are typically of the order $1\times 10^{-17}$ erg s$^{-1}$ cm$^{-2}$ \AA$^{-1}$. 

Two independent spectral reduction pipelines assign redshifts and classifications to these spectra and measures fluxes of several major emission lines. The {\verb spectro1d } pipeline \citep{stou02,subb02}, fits Gaussian profiles to emission lines to determine emission line redshifts, and the {\verb specBS } pipeline, written by D. Schlegel, carries out $\chi^2$ fits of spectra to templates in wavelength space. For 98\% of spectra, the measured redshifts from the two pipelines agree within 300 km s$^{-1}$ for galaxies, and 3000 km s$^{-1}$ for quasars \citep{adel08}; but some type 2 quasars are among the 2\% of the discrepant objects and we must treat them separately (\S\ref{subsec:sdss_selec}). 

\subsection{SDSS Spectroscopic Target Selection} 
\label{subsec:sdss_spec}

Spectroscopic target selection in the SDSS is based on a combination of photometric properties, such as magnitudes, colors, and morphologies, and in some cases, radio and X-ray properties. In this section, we describe the subset of target algorithms from the `Main survey' and the `Special Southern Survey' which are important in the selection of type 2 quasars, and subsequently, for our calculation of their luminosity function in \S\ref{sec:type2_lf}.

The Main survey constitutes close to complete samples of galaxies and quasars, which comprise roughly 75\% of all SDSS spectra. The Galaxy algorithm targets resolved sources down to a limiting Petrosian magnitude of $r=17.77$ \citep{stra02} after correction for Galactic extinction following \citet{schl98}. These objects represent about 62\% of the objects in the Main survey. The Luminous Red Galaxy (LRG) algorithm targets sources based on the distinctive colors of LRGs, down to a Petrosian magnitude of $r=19.5$ \citep{eise01}, representing 9\% of the Main survey. The Low-z QSO algorithm targets mostly UV excess sources down to $i_{\rm PSF}=19.1$, and the High-z QSO algorithm targets point sources with the colors of high-redshift ($z>3$) quasars down to $i_{\rm PSF}=20.2$ \citep{rich02}. Taken together, quasar candidates represent 13\% of the Main survey. 

Other Main survey target algorithms are assigned spectroscopic fibers only after the galaxy and quasar targets have been allocated, so they do not produce complete samples. Objects with unusual colors, radio emission detected by the FIRST survey  (`Serendipity FIRST'; \citealt{beck95}), and X-ray emission detected by the ROSAT survey (`ROSAT'; \citealt{voge99}, see also \citealt{ande07}) are selected down to a fiber magnitude of $i=20.5$ \citep{stou02}. These so-called `serendipity' targets constitute about 5\% of the Main survey.

Another 10\% of spectra are taken in the Equatorial Stripe of the Southern Galactic Cap, covering $\sim 300$ deg$^2$, as part of the Special Southern survey \citep{adel06}. Spectroscopic targeting algorithms for this survey are somewhat different from those in the Main survey. Those important for the selection of type 2 quasars are: (i) modified versions of the Main survey algorithms, which target galaxies and quasars to limiting magnitudes $\lesssim$ 1 mag fainter, (ii) the `faint quasars' algorithm, which is a modified version of the Main Low-z QSO algorithm with looser color criteria, and (iii) the `photoz' algorithm, which targets sources in the blue end of the normal galaxy distribution down to a Petrosian $r$-band magnitude of 19.5. These algorithms are exploratory in nature and have changed several times over the course of the survey. For this reason, we do not use objects targeted by these ``incomplete'' algorithms in the calculation of the luminosity function. 

\subsection{Spectroscopic Selection of Type 2 Quasars}
\label{subsec:sdss_selec}

Following Paper I, we select Type 2 quasars as objects with narrow emission lines without underlying broad components, and with line ratios characteristic of non-stellar ionizing radiation. We searched the entire SDSS spectroscopic database as of 2006 July, which contains 1.08 $\times 10^6$ spectra from 1770 plates, before accounting for duplicates. This corresponds to $\sim$ 80\% of the DR6 spectroscopic database \citep{adel08}, and is three times larger than the one used in Paper I. 

The type 2 quasar selection was performed using the spectrophotometric calibration used in the First Data Release \citep{abaz03}. The Second Data Release paper \citep{abaz04} describes a substantial improvement in the spectrophotometric calibration algorithm, which affects the ratios of line strengths between the red and the blue end of the spectrum.  An additional modification in the algorithm was incorporated in the Sixth Data Release \citep{adel08}, whereby spectra are calibrated relative to PSF rather than fiber magnitudes of the standard stars measured on each plate.  Thus, the flux scale used in our luminosity calculations (\S\ref{subsec:type2_o3} and \S\ref{subsec:type1_o3}) is on average 38\% higher than what was used in the initial selection.  These changes in calibration cause some minor incompleteness at the faint end of the luminosity function, but very few objects change their classification from AGN to star forming galaxy (Eqs.~1--3 below) due to these changes.

Our automated selection algorithm applies a series of constraints on the data: a redshift maximum, a spectroscopic signal-to-noise ratio (S/N) minimum, an \oiiil {} luminosity minimum, and a set of emission line ratio cuts. The resulting $\sim 4000$ spectra are then visually inspected and fit for broad components in H$\beta$ and H$\alpha$ for final selection. Redshifts are also checked for accuracy.

\begin{figure}
\epsscale{0.9}
\plotone{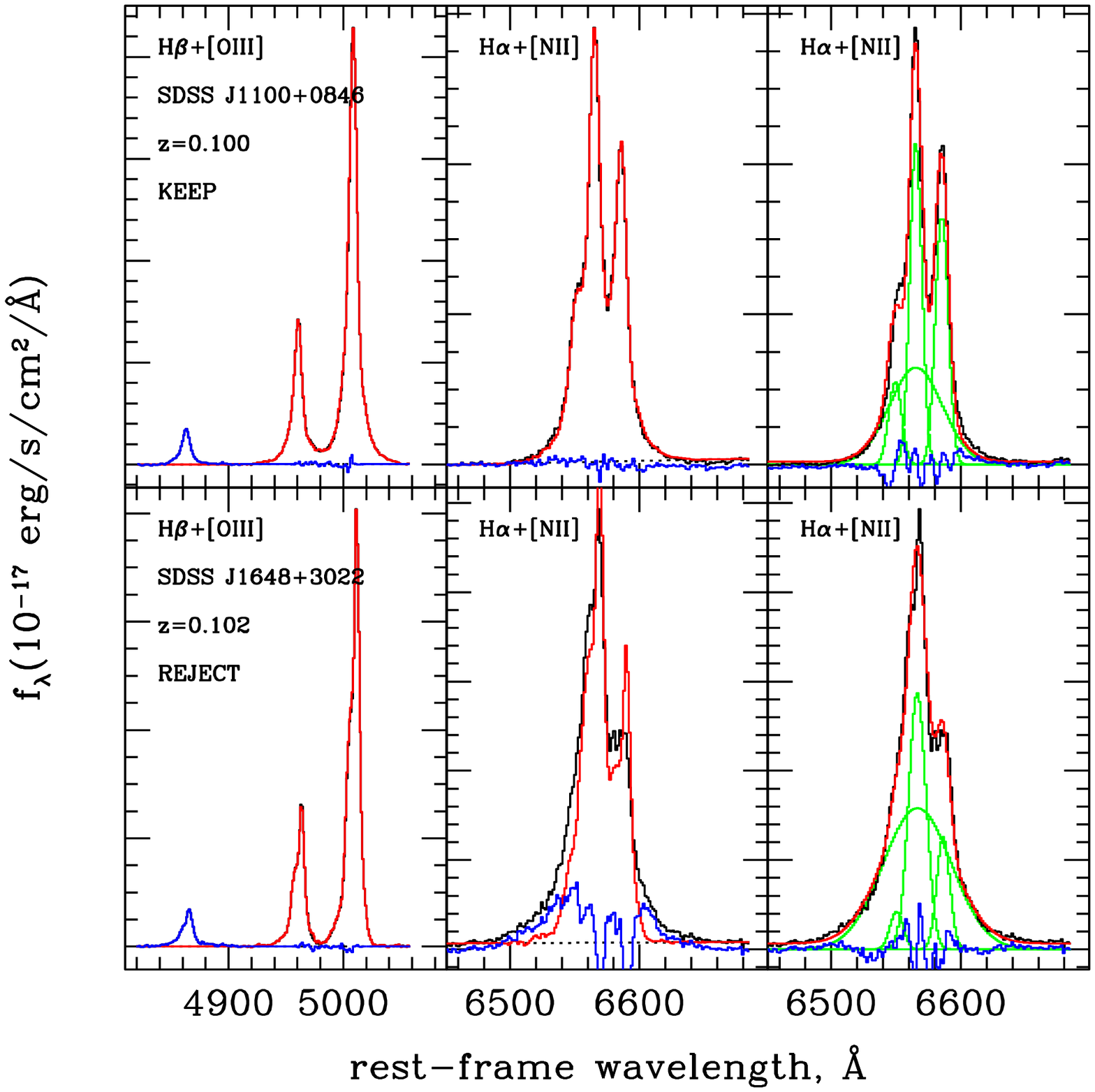}
\figcaption{Examples of our non-parametric fitting procedure to determine whether there is a broad component in H$\alpha$ line. From left to right, the three panels for each object are: (i) the H$\beta$+[OIII] complex and model fits to the blue wing of [OIII]4959 and the red wing of [OIII]5007; (ii) fit of the H$\alpha$+[NII] complex, using the [OIII]4959, 5007 line profile for all three lines; (iii) the best 4-Gaussian fit to the H$\alpha$+[NII] complex. In each panel, the original spectrum is in black, the model is in red, residuals are in blue and the four Gaussian components are in green. Although for both objects the 4-Gaussian fit to the H$\alpha$+[NII] complex is statistically preferred to the 3-Gaussian fit, in SDSS J1100+0846 (upper panel) this complex is well-fit with a blend of lines shaped like [OIII], so we keep this object in our sample. In SDSS J1648+3022 (lower panel), the complex is significantly broader than the blend of [OIII] lines, as indicated by the fit residuals, so we classify this object as a broad-line AGN and exclude it from our sample. \label{pic_halpha}}
\end{figure}

The major differences in our selection algorithm from that in Paper I are: (i) we now impose a luminosity cut, \loiii {} $\ge 10^{8.3} L_\odot$, because selection based on emission line ratios becomes incomplete and noisy at low luminosities; (ii) we consider all objects with redshifts $z<0.83$ (our previous sample was at redshifts $0.3<z<0.83$, complemented by the $z<0.3$ AGN samples by \citealt{hao05a} and \citealt{kauf03}); and (iii) we use an improved algorithm for identifying weak broad components in H$\alpha$ and H$\beta$ (Fig.~\ref{pic_halpha} and text below). 

We restrict the selection to objects with redshifts $z<0.83$ so that the \oiiil {} line (the strongest expected emission line) is present in all spectra. In order to select emission line objects, we require the rest-frame equivalent width of \oiiil {} to be $>4$\AA. In addition, the signal-to-noise ratio must be $\ge 7.5$, where the signal is the flux density in the seventh brightest pixel over the entire spectroscopic range (3800--9200 \AA; about 3840 pixels) and the noise is the median estimated flux error per pixel over all pixels. This unconventional criterion allows retention of objects with weak continua but strong narrow emission lines, while rejecting continuum-dominated sources with low signal-to-noise ratio (Paper I).

When the redshift and classification of the two reduction pipelines \verb spectro1d {} and \verb specBS {} (described in $\S$\ref{subsec:sdss_data}) agree, we apply different emission line criteria depending on the redshift of the object. For objects with redshifts $z<0.36$, both H$\beta$+\oiii 4959,5007 and H$\alpha$+[NII] 6548,6583 line complexes are covered by the spectroscopic data, and the classical emission line diagnostic diagrams \citep{bald81, veil87, oste89} are used to distinguish between a stellar and an AGN ionizing continuum. The MgII 2799 emission doublet is either not covered by the spectrum or falls into the UV where the spectra typically have low signal-to-noise ratio, so this line is not used for this set of spectra. We use line diagnostic criteria of the form suggested by \citet{kewl01} to distinguish type 2 quasars from star-forming galaxies and narrow-line AGN. We require the ratio of luminosities $\cal{R} \equiv {\rm [OIII]5007}/{\rm H\beta}$ to satisfy either:
\beq
\log\left(\cal{R}\right)>\frac{0.61}{\log({\rm [NII]6583/H\alpha})-0.47}+1.19\label{diag1}
\eeq
or
\beq
\log\left(\cal{R}\right)>\frac{0.72}{\log({\rm [SII]/H\alpha})-0.32}+1.30,\label{diag2}
\eeq
where [SII] refers to the combined luminosity of the doublet {\rm [SII]6716,6730}.

For objects with redshifts $0.36\le z<0.83$, the H$\alpha$+[NII] line complex is not covered by the SDSS spectra, so the classical diagnostic diagrams cannot be used. We adopt the \oiiil/H$\beta$ ratio requirement in reduced form:
\beq
\log \left(\cal{R}\right) > 0.3\mbox{, if H$\beta$ is detected with S/N $>$ 3} \label{eq:c2}
\eeq
or we require that H$\beta$ is undetected, while \oiiil {} is detected. In addition, for $z>0.6$, the full width at half maximum (FWHM) of MgII 2799 is required to be $<2000$ km/s. 

In the $2$\% of cases for which the two spectroscopic reduction pipelines do not agree on redshift or classification, we find the emission line closest to the expected position of \oiii 5007, given either of the two redshifts. If the line satisfies the appropriate equivalent width and luminosity criteria above, we retain the object as a candidate.

These selection criteria are designed to be maximally inclusive. In particular, the line diagnostic criteria given in Eqs.~\ref{diag1} and \ref{diag2} are applied with the `OR' operator, in case some of the lines are not measured properly. At redshifts $z>0.36$, only a very mild line ratio cut (Eq.~\ref{eq:c2}) is imposed. Apart from the MgII 2799 width criterion on $z>0.6$ objects, no criteria to explicitly reject broad-line AGN are imposed at this selection stage, since the weak H$\beta$ lines common in type 2 AGN are often poorly measured by the spectroscopic pipelines. Nevertheless, the vast majority of type 1 AGN are easily rejected by the automated procedure; because of their broad H$\alpha$ and H$\beta$, they tend to have small values of $\log$([OIII]5007/H$\beta$), $\log({\rm [NII]6583/H\alpha})$ and $\log({\rm [SII]6716,6730/H\alpha})$, making these objects fail the line diagnostic criteria given in Eqs.~\ref{diag1} and \ref{diag2}. This automated procedure selected $\sim$ 4,000 objects for visual inspection.

At the visual inspection stage, we pursue two goals: (i) to remove objects with broad components in their permitted emission lines (the major contaminant at $z<0.36$); (ii) to reject star-forming galaxies (the major contaminant at $z\ge 0.36$, since at $z<0.36$ the diagnostic diagrams are robust). We consider the [NeV] 3346, 3426 emission lines to be unambiguous signs of an underlying AGN continuum \citep{grov04,naga06}, so if either one of these lines is detected (if a single line is detected, it is usually [NeV] 3426), the object is considered an AGN. If neither of the [NeV] lines is detected, we use the criterion FWHM([OIII]5007)$>$400 km/s as a sign of AGN activity \citep{zaka03, hao05a}. Since the [NeV] lines are weak, our selection is more robust at high luminosities, where these lines are more likely to be detected.  

Objects identified as AGN are then checked for the presence of broad lines. A broad component in the H$\beta$ line can be rejected by examining the difference in $\chi^2$ of a single Gaussian and a double Gaussian fit to this line, similar to \citet{hao05a}. However, identifying weak broad components in H$\alpha$ is complicated by the non-Gaussianity of emission lines and blending with [NII] 6548, 6583. Thus, we do not automatically reject objects for which a four-Gaussian fit is statistically preferred to a three-Gaussian fit to the H$\alpha+$[NII] line complex. Instead, we use the non-parametric line fitting procedure illustrated in Fig.~\ref{pic_halpha}. First, we derive narrow line profiles from the [OIII]4959, 5007 emission lines, and then fit the H$\alpha$+[NII] complex assuming that all three lines have the same profile (with the ratio [NII]6583/[NII]6548 fixed to 3). We visually examine the residuals, and if the non-parametric fit can reproduce the width of the complex, we retain the object in our sample.

\begin{figure}
\epsscale{0.9}
\epsfig{file=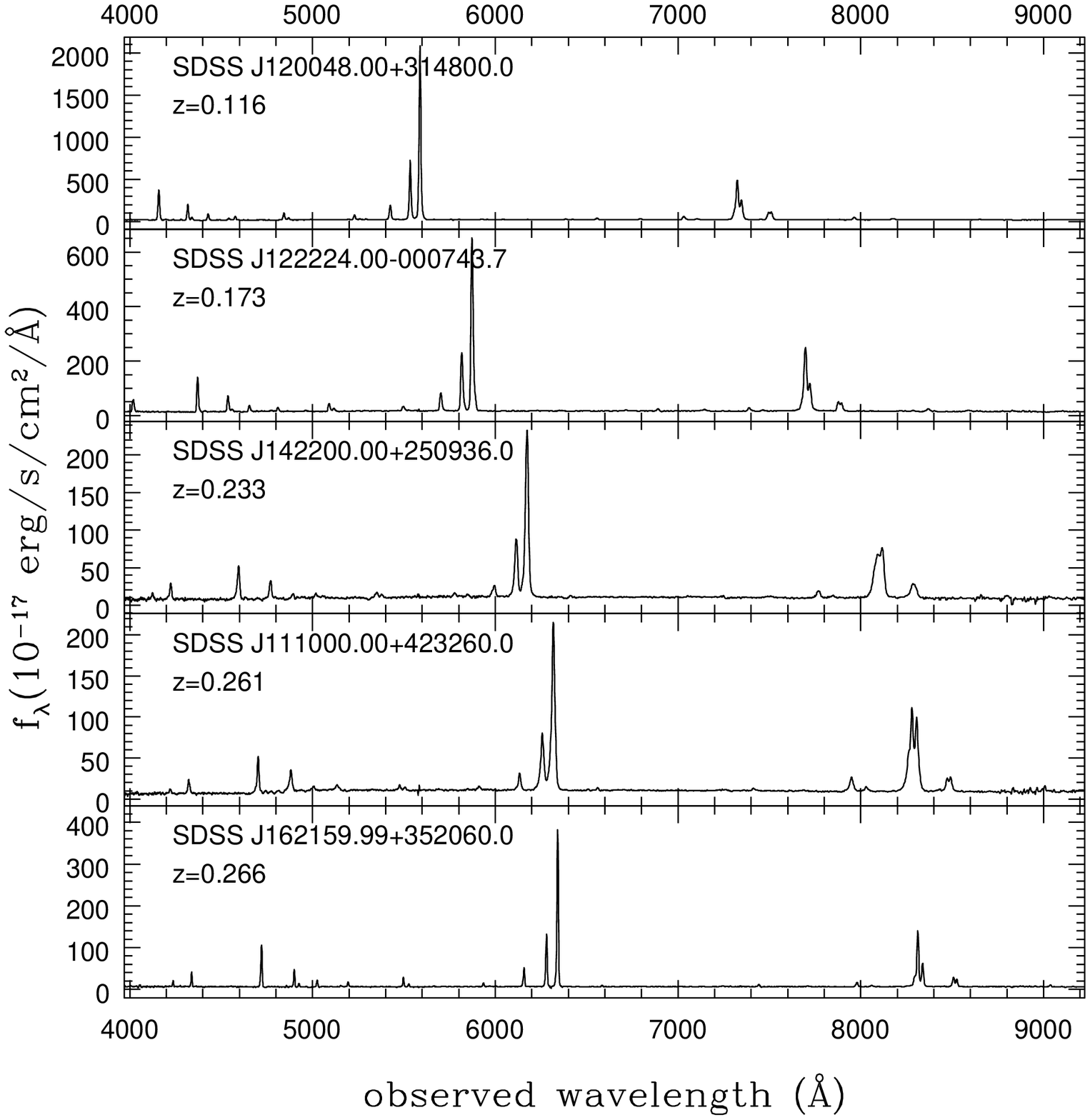,width=0.3\linewidth}
\epsfig{file=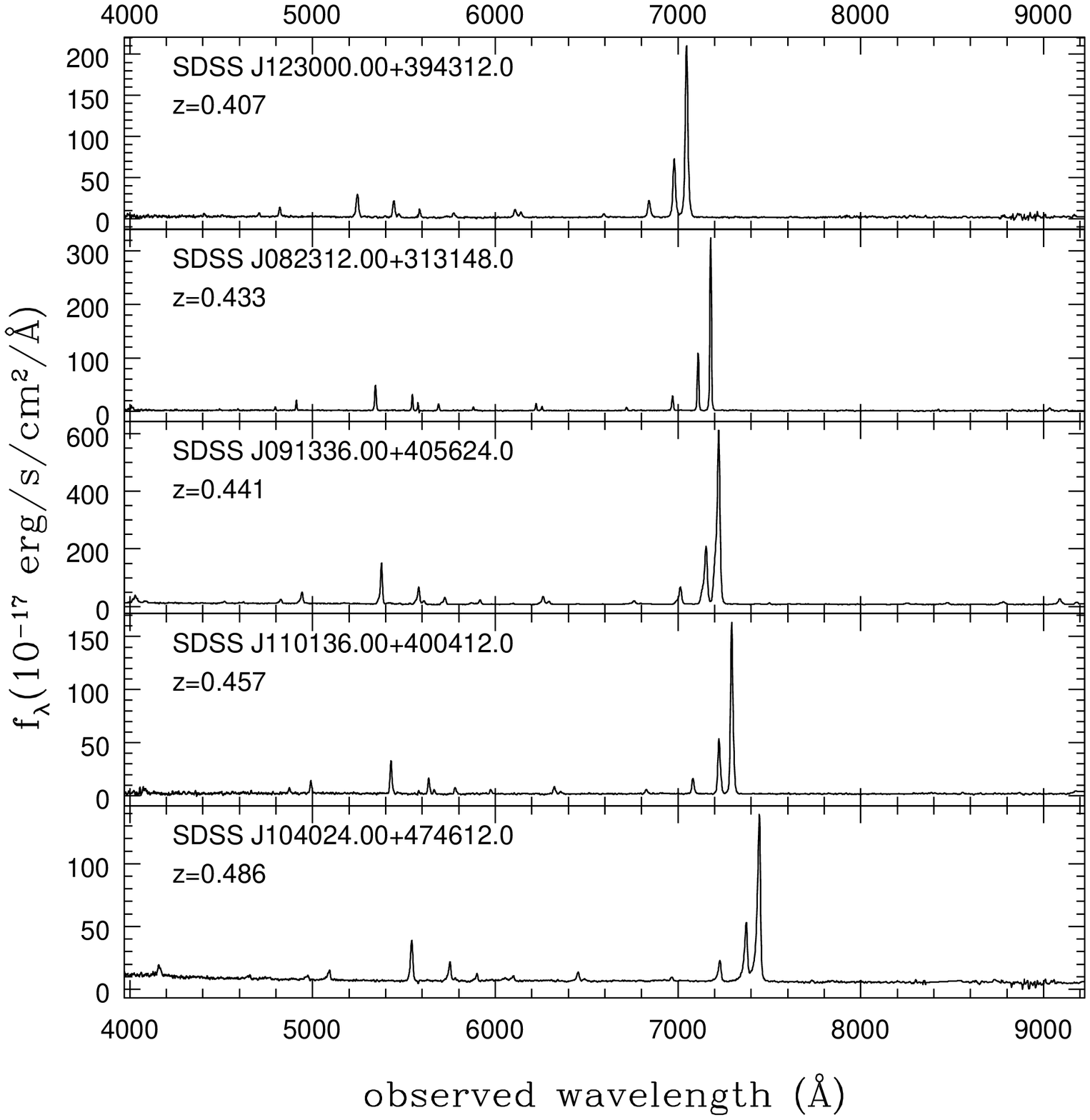,width=0.3\linewidth}
\epsfig{file=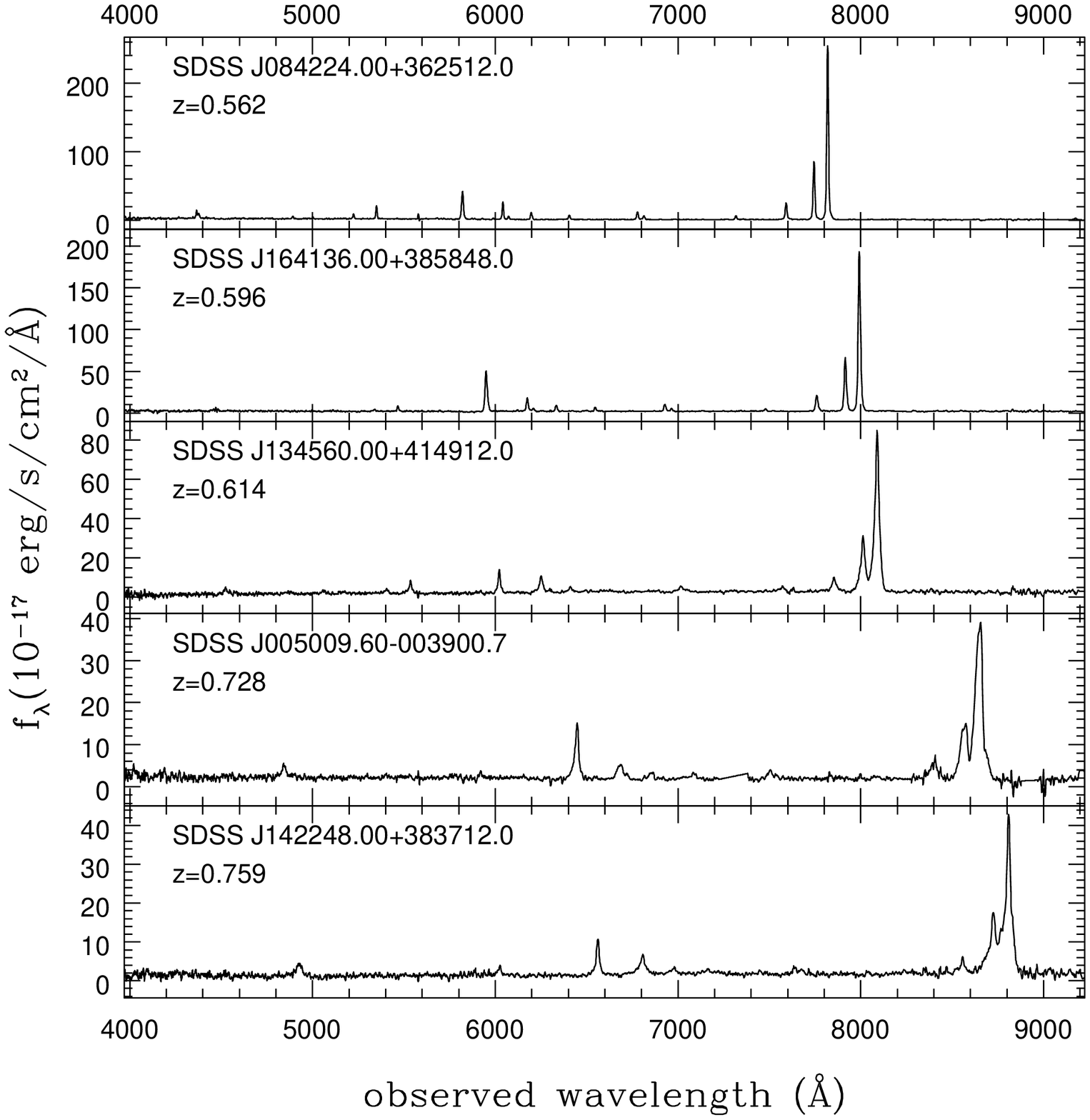,width=0.3\linewidth}
\figcaption{Example spectra of the highest luminosity type 2 quasars in the redshift range \za, \zb, and \zc {} smoothed by 5 pixels for display purposes. Objects in the three panels have \loiii $>1.8 \times 10^9$, $5.0 \times 10^9$, and $9.6 \times 10^9$ $L_\odot$, respectively. The strongest emission line is [OIII]5007. \label{fig:qso2_spec_z0}}
\end{figure}

This selection procedure results in the sample of 887 type 2 quasars given in Table \ref{tab:qso2}. Figure \ref{fig:qso2_spec_z0} shows sample spectra of high \oiiil {} luminosity objects in our sample, in order of increasing redshift. More than $90\%$ of the type 2 quasar candidates that were selected in Paper I are recovered by the automated selection procedure described above, so we estimate that roughly the same percentage of type 2 quasars in the SDSS database are successfully selected. The spectra of those objects from Paper I that we did not recover tend to be of low signal-to-noise ratio or have ambiguous classification.

\subsection{Measurement of \oiiil {} Luminosities}
\label{subsec:type2_o3}

We measure \oiiil {} line luminosities from spectra calibrated with the most recent spectrophotometric calibration algorithm (\citealt{adel08}, also see beginning of \S\ref{subsec:sdss_selec}). We do not correct the measured line luminosities for dust extinction because of the large uncertainties involved in such a correction; we discuss how this effect might affect our results in \S\ref{subsec:reddening}. We fit the \oiii 4959,5007 lines with two Gaussians of the same width and a fixed $1:3$ amplitude ratio, plus a linear continuum, over the wavelength range 4860--5060{\AA}. We note that \feii {} emission tends to be negligible in type 2 quasars (unlike in type 1 quasars, for which we carefully subtract \feii {} emission before measuring \oiiil {}  luminosities, as described in $\S$\ref{subsec:type1_o3}).

To check the fits, we also obtain a non-parametric measure of the \oiiil {} line luminosity by integrating the detected flux density and subtracting the continuum contribution. As Fig.~\ref{pic_halpha} shows, this line is often significantly non-Gaussian, especially at its base. Asymmetries in line profiles, thought to be caused by radial outflows in the narrow-line region, have been observed in many AGN \citep{heck81,whit85}. We find that the non-parametric measure of the luminosity is systematically larger, by 5\% on average, than the Gaussian measure; this indicates the presence of non-Gaussian wings in the \oiiil {} profiles of objects in our sample.

Both Gaussian and non-parametric measures are given in Table \ref{tab:qso2}. In the rest of the paper, \loiii {} refers to luminosities measured from Gaussian fits. About 30\% of the sample (257 objects) have \loiii {} $>10^{9.5} L_\odot$. As we explained in \S\ref{subsec:sdss_selec}, the luminosity cut we imposed in our selection, \loiii {} $>10^{8.3} L_\odot$, was applied to luminosities calibrated using a flux scale that is $\sim 38\%$ lower than the one we use for these line measurements. Due to this change in spectrophotometric calibration, there are 149 objects with measured \oiiil {} luminosities below $10^{8.3} L_\odot$; of these, around 80\% are within 30\% of this value. 

\section{Type 2 Quasar Luminosity Function}
\label{sec:type2_lf}

In this section, we derive the \oiiil {} luminosity function from our sample of type 2 quasars. Since type 2 quasars are heavily obscured, the optical continuum luminosity is by definition a poor indicator of their bolometric luminosity. The \oiiil {} emission line offers a plausible alternative, since it is clearly detected in both type 1 and type 2 AGN, and is observed to correlate with continuum luminosity for type 1 sources (see discussion in \S\ref{subsec:type1_rich} and \S\ref{subsec:disc_o3}). \citet{hao05b} and \citet{simp05} derived the \oiiil {} luminosity function from a sample of type 2 AGN with redshifts $z \lesssim 0.3$ and \oiiil {} luminosities up to $\sim 10^{8.6} L_\odot$ and $10^{8.9} L_\odot$, respectively. Our sample probes both larger redshifts and higher luminosities, up to $z = 0.83$ and \loiii {} $\sim 10^{10} L_\odot$. To minimize the systematic effect of redshift evolution within the sample, we present the LF for three ranges in redshift: \za, \zb {} and \zc.

The type 2 quasar LF we derive is a lower limit to the true LF for several reasons. First, obscured quasars for which the narrow emission-line region is also obscured would not meet our selection criteria, so they are not included in our sample \citep{rigb06}. Second, there may be objects that fall outside the regions in the multi-dimensional parameter space of colors, magnitudes and morphologies, which the SDSS spectroscopic target algorithms are designed to target, so they are also not included in our sample. Third, we use line luminosities that are not corrected for dust extinction as it is unclear how to do so consistently (see \S\ref{subsec:reddening}). Correction for line extinction would shift the LF toward higher luminosities and would yield a higher space density at any given luminosity.

\begin{figure}
\plotone{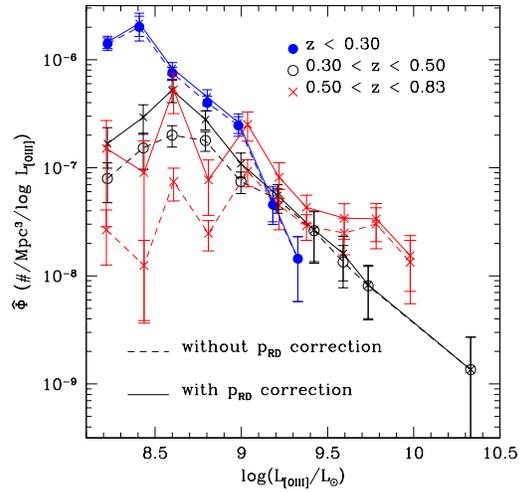}
\figcaption{\oiiil {} luminosity function of type 2 quasars (lower limits) for three redshift ranges,  with and without correction for the probability of radio detection (see \S\ref{subsec:vvmax_alpha};  solid and dashed curves, respectively). The maximum volume is calculated using the selection criteria of the various SDSS spectroscopic target selection algorithms (see \S\ref{subsec:type2_selec}). Included in the calculation are 871 objects targeted by the top four Main survey algorithms (Galaxy, Low-z QSO, High-z QSO, and Serendipity FIRST).\label{fig:phitable_qso2_zbin}}
\end{figure}

In \S\ref{subsec:type2_invvmax} and \S\ref{subsec:type2_selec}, we discuss the calculation of the LF, and in \S\ref{subsec:type2_results}, we present the results and compare it with previous work. The derived type 2 quasar LF is shown in Fig.~\ref{fig:phitable_qso2_zbin}. In \S\ref{subsec:type2_missed}, we discuss the efficiency of our selection of type 2 quasars and its implications to the interpretation of the derived LF.

\subsection{\invvmax {} Luminosity Function}
\label{subsec:type2_invvmax}

To derive the luminosity function, we use the \invvmax {} estimator \citep{schm68}, in which the contribution of each object to the luminosity function is weighted by its available volume \vmax. To calculate \vmax, we need to determine how each object had been selected for spectroscopy. The SDSS spectroscopic target selection algorithms that are important in selecting type 2 quasars are described in \S\ref{subsec:sdss_spec}. They are listed in Table \ref{tab:target} together with the number of objects targeted by each algorithm. An object can be targeted by multiple target algorithms, so these numbers do not sum to the total number of objects in the catalog. As noted in \S\ref{subsec:sdss_spec}, target algorithms of the Special Southern survey have changed over time \citep{adel06}, so we do not include objects from this survey in the calculation of the LF. Furthermore, we include only objects targeted by any of the four primary Main survey algorithms (Galaxy, Low-z QSO, High-z QSO and Serendipity FIRST), which together represent 83\% of the full catalog (740 out of 887 objects). Fig.~\ref{fig:lfcat} shows the distribution of \oiiil {} luminosities and redshifts of objects targeted by these primary algorithms.

In calculating \vmax, the line luminosity and S/N criteria we have applied in $\S$\ref{subsec:sdss_selec} are not important for the luminosity and redshift ranges of interest here, since the \oiiil {} line is strongly detected in all objects in our sample. We therefore focus on the selection criteria from each spectroscopic target selection algorithm. For our purposes, \vmax {} is given by the comoving volume over which an object would be selected by any of the four primary target algorithms:
\begin{equation} 
V_{\rm max} = \frac{\Omega}{4\pi} \int_{z_1}^{z_2} \Theta(z) \frac{dV_{\rm c}}{dz}(z){}  dz,\label{eq:vmax}
\end{equation}
where $\Omega$ is the effective survey area, $z_1$ and $z_2$ are the edges of the redshift range for which we are calculating the luminosity function, $(dV_{\rm c}/dz) {} dz$ is the comoving volume element in the redshift interval $dz$, and $\Theta(z)$ is a generalized step function that is equal to zero if the object is not selected by any of the four algorithms we consider, and nonzero otherwise. We describe in detail how we calculate $\Theta(z)$ in \S\ref{subsec:type2_selec}. 

The effective survey area $\Omega$ is given by the area covered by the Main survey spectroscopic plates from which we have selected objects. To calculate $\Omega$, we define areas of intersection of plates and tiling rectangles and take the area covered by their union (see Appendix B of \citealt{shen07} and references therein for details). This calculation yields $\Omega \approx 6293$ deg$^2$. Since some post-DR5 plates are included in our parent sample, this area is slightly bigger than the DR5 spectroscopic footprint area of 5740 deg$^2$ \citep{adel07}.

The \invvmax {} luminosity function and its uncertainty are given by
\begin{equation} 
\Phi(\overline{L}_k) = \frac{1}{(\Delta L)_k} \sum_{j=1}^{N_k} \left(\frac{1}{V_{{\rm max},j}}\right) \label{eq:phi}
\end{equation}
and 
\begin{equation}
\sigma(\Phi) =\frac{1}{(\Delta L)_k} \left[ \sum_{j=1}^{N_k} \left(\frac{1}{V_{{\rm max},j}}\right)^2 \right]^{1/2}, \label{eq:phierr}
\end{equation}
where $\overline{L}_k$ is the mean \oiiil {} luminosity of objects in the $k$th luminosity bin ($L$,{} $L+(\Delta L)_k$), and the sum is over the $N_k$ objects in that bin. Following common practice, we present our results in terms of the number of quasars per unit volume per logarithmic luminosity interval, denoted by $\hat{\Phi}(L) = (L/\log_{10} e) \Phi(L)$.

\subsection{Calculation of the Selection Function $\Theta(z)$}
\label{subsec:type2_selec}

In this section, we calculate the function $\Theta(z)$, which appears in the expression for \vmax {} given above (Eq.~\ref{eq:vmax}). We determine whether an object, with redshift $z_{\rm obs}$, would be selected by any of the four primary target algorithms, if it were placed at some other redshift $z$, based on its intrinsic spectral energy distribution (SED). For this object, $\Theta(z)$ is given by a generalized step function: 
\beq \label{eq:thetaz}
\Theta(z) =  \left\{ \begin{array}{ll} 
1 & \mbox{, if selected by any of \textit{G}, \textit{L}, or \textit{H}}\\
\alpha & \mbox{, if selected solely by \textit{S}}\\
0 & \mbox{, if not selected by \textit{G}, \textit{L}, \textit{H}, or \textit{S}}
 \end{array} \right.
\eeq
where \textit{G}, \textit{L}, \textit{H}, and \textit{S} stand for the Galaxy, Low-z QSO, High-z QSO, and Serendipity FIRST algorithms, respectively, and $\alpha$ is a numerical factor (smaller than 1) that weights the contribution of the radio-selected objects to the LF. We discuss the determination of $\alpha$ in \S\ref{subsec:vvmax_alpha}. In practice, we determine $\Theta(z)$ in steps of redshift, from $z=0.1$ to $0.8$, separated by $\Delta z = 0.01$, and we approximate the integral in Eq.~\ref{eq:vmax} as a sum over these redshift slices. 

To determine $\Theta(z)$ for a given object, we calculate what its observed optical and/or radio properties would be if it were placed at redshift $z$, and apply the relevant selection criteria for each algorithm. For all objects, we apply k-corrections \citep{sand61} based on their observed SDSS spectra, which take into account both the shape of the SED and the shifting of emission lines in and out of bandpasses. The latter effect is important because objects in our sample have equivalent widths as large as 1,400\AA {} (Paper I).  To calculate k-corrections for the SDSS $u$ and $z$ bands, we extrapolate the observed spectrum outside $\sim$ 3,000$-$10,000 \AA {} using a constant flux density; our results are not sensitive to this extrapolation. In the following subsections, we describe details in the procedure for determining $\Theta(z)$ that are specific to each target algorithm.

\subsubsection{Main Galaxy Target Algorithm}
\label{subsec:vvmax_gal}

The Galaxy algorithm targets objects that have Petrosian magnitude $r<17.77$ mag, $r$ band Petrosian half-light surface brightness $\mu_{50} \le 24.5$ mag arcsec$^{-2}$, and satisfy a star/galaxy separation cut and a fiber magnitude cut \citep{stra02}. 

Since surface brightness changes only by a factor of $\sim 3$ throughout the redshift range probed by the Galaxy algorithm ($z \lesssim 0.3$; Fig.~\ref{fig:lfcat}), and no object in the sample is close to the star/galaxy separation cut, the magnitude cut alone determines whether an object is selected. We consider an object to be selected by this algorithm if its scaled and k-corrected Petrosian $r$-band magnitude is brighter than 17.77 mag. We assume for simplicity that the spectral shape within the Petrosian aperture is the same as within the spectroscopic fiber.

\subsubsection{Low-z QSO and High-z QSO Target Algorithms}
\label{subsec:vvmax_qso}

Low-z QSO targets must have $15.0 < i_{\rm PSF} < 19.1$ and satisfy various color criteria, i.e., they must be outliers from the $ugri$ stellar locus, have non-galaxy colors if they are extended, and not occupy any of several exclusion regions in color space designed to eliminate white dwarfs, A-type stars, and other contaminants \citep{rich02}. All UV excess sources ($u-g < 0.6$) that are not in the white dwarf exclusion region and satisfy the magnitude limits are also targeted.

The High-z QSO target algorithm is designed to recover quasars at redshifts beyond $\sim 3.0$, as the Lyman break moves across the photometric bands with increasing redshift. High-z QSO targets must be point-like sources with $15.0<i_{\rm PSF}<20.2$ \citep{rich02}. They must also be outliers in the $griz$ stellar locus, or occupy certain regions in the color space where high-redshift quasars are expected to lie.

To determine whether an object would be selected by these algorithms at a redshift $z$, we run the scaled and k-corrected PSF magnitudes and errors in all bands through the final version of the QSO target selection code \verb v3_1_0 {} \citep{rich02}. 

\subsubsection{Serendipity FIRST Target Algorithm}
\label{subsec:vvmax_sf}

Serendipity FIRST targets are sources that have fiber magnitudes $14.0 < g,r,i < 20.5$ and have counterparts in the FIRST catalog within 2\arcsec\ of the optical position. Radio sources are included in the FIRST survey catalog if they have peak 20 cm radio flux density $S_{20 {\rm cm}} > 5\sigma + 0.25$ mJy/beam and $S_{20 {\rm cm}} > 1$ mJy/beam, where $\sigma$ is the local RMS noise in the field \citep{beck95}.

To calculate the fiber magnitude as a function of redshift, we use the azimuthally-averaged radial surface brightness profile (following Appendix B of \citealt{stra02}) to calculate the light that would fall into the aperture at each redshift, neglecting the effect of seeing. To calculate the radio flux density as a function of redshift, we assume a power-law radio spectrum $F_\nu = A \nu^\beta$ with $\beta = -0.5$, following \citet{zaka04}. Most of our objects are point radio sources, so we approximate the redshift scaling of the peak flux density $S_{20 {\rm cm}}$ to be the same as the total flux $F_\nu \propto D_{\rm L}^{-2}(z) (1+z)^{1+\beta}$, where $D_{\rm L}(z)$ is the luminosity distance at redshift $z$. 

Since most objects targeted by the Serendipity FIRST algorithm were selected in the $i$-band and because the volume corresponding to the bright limit is negligibly small, we consider an object to be selected by this algorithm if it satisfies both the $i$-band fiber magnitude faint limit and the radio flux limits.

\subsubsection{Probability of Radio Detection}
\label{subsec:vvmax_alpha}

About 30\% of the objects included in our LF calculation are selected solely by the Serendipity FIRST algorithm. These objects do not have the optical morphologies or colors that would allow them to be selected by the three other algorithms we consider. For this set of objects, we consider two definitions for the selection function $\Theta(z) = \alpha(z,L_{\rm [OIII]})$ depending on whether an explicit correction for the probability of radio detection $\pzl$ is applied, i.e.,
\begin{equation} \label{eq:alpha}
\alpha(z,L_{\rm [OIII]}) = \left\{ \begin{array}{l}
f_{\rm obs} \\
f_{\rm obs} \times \pzl
\end{array} \right.
\end{equation}
Here, $f_{\rm obs}=10,480/25,307=0.414$ accounts for the incompleteness in the  spectroscopic observations of these ``serendipity'' targets, and is simply the fraction of targets with observed spectra. The factor $\pzl$ accounts for the contribution of radio-weak type 2 quasars with optical SEDs similar to the radio-selected sources (and therefore would not be selected by the other algorithms), but which are too radio-faint to have been detected by the FIRST survey. We define it to be the probability that an object with a given redshift and \oiiil {} luminosity has a FIRST catalog match \citep{beck95} within 2\arcsec\ of its optical position. In \S\ref{subsec:type2_results}, we show the type 2 quasar LF calculated both with and without this correction.

The distribution of radio luminosities in AGN and the relation, if any, between optical and radio properties are the subjects of much debate (\citealt{ivez04} and references therein). Here, we determine $\pzl$ from our data alone. We use the radio properties of type 2 quasars in our sample that are selected by at least one non-radio method (539 objects) to predict how many objects the Serendipity FIRST algorithm has missed because they fall below the flux limit of the FIRST survey. In the following calculation, we assume that the distribution of radio luminosities of a population of type 2 quasars is independent of their optical colors and morphologies, but may depend on their \oiiil {} luminosities and redshifts. 

We aim to find a simple analytic functional form for $\pzl$ that best fits the available data. We define a statistic analogous to $\chi^2$ to assess the goodness of the fit. For a bin $j$ in the space of redshift and \oiiil {} luminosity, which contains $N$ objects with redshifts and luminosities $z_i$ and $L_i$ ($i=1...N$) of which $n$ are detected by FIRST, this statistic is given by
\beq\label{eq_u}
u=\frac{N(f-\bar{p})^2}{\bar{p}(1-\bar{p})},
\eeq
where $f=n/N$ and $\bar{p}=\sum_i \prad(z_i,L_i)/N$. In the limit of large $N$ and small bins (so that $\prad \simeq$ constant within each bin), we show in the Appendix that $u$ is distributed as $\chi^2$ with one degree of freedom. Therefore, the central limit theorem can be applied to the sum of the values of $u$ over all bins, and this value gives the probability that the given form of $\pzl$ fits the data. Importantly, this statistic is independent of the distribution of objects in the $z$-\loiii {} plane; this distribution is strongly affected by selection effects. 

\begin{figure}
\epsscale{0.9}
\plotone{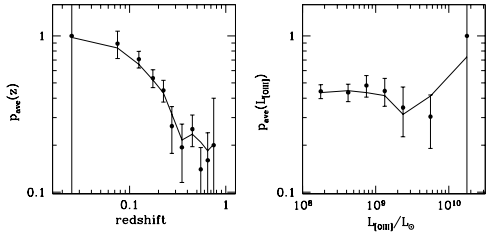}
\figcaption{Comparison of the observed probability of FIRST detections with the best-fitting function for $\pzl$ {} (Eq.~\ref{eq:pzl}) for non-radio-selected type 2 quasars. Left: in each redshift range $z_1\le z<z_2$, the circle shows the value $n/N$, where $n$ is the number of radio detected objects in this bin and $N$ is the total number of objects. The Poisson error bars reflect the total number of objects contributing to each bin. The solid line connects values $p_{ave}(z)=\sum_i \prad(z_i,L_i)/N$, where the summation is over all objects in this bin. Right: same, but for luminosity bins instead of redshift bins. Such representation allows us to properly take into account the distribution of objects in the $z$-\loiii {} plane.\label{fig:pzl}}
\end{figure}

We expect that the probability of radio detection decreases with redshift, since more distant objects are dimmer, and increases with luminosity. We tried several functional forms for $\prad$ with this in mind and varied their parameters to minimize the sum of $u$ over all bins. Our best-fit function is given by
\begin{equation} \label{eq:pzl}
\prad(z,L)=\frac{1}{1+[z/(0.15+0.1(\log L/L_{\odot}-8.0)^2)]^2}.
\end{equation}
The comparison of observed FIRST detection rates with those calculated using this function is shown in Fig.~\ref{fig:pzl}. By visually examining the agreement between the observed detection rates and $\prad$ (using slices and projections in $z$-\loiii {} plane) we estimate that $\prad$ is determined to better than $\pm 10\%$ over most of the parameter space. Since $\pzl \sim 1$ at low redshifts, as well as at high \oiiil {} luminosities, the correction to the LF from this factor is not important in these regimes. In \S\ref{subsec:test_radio}, we test whether the radio-detection rates of type 2 quasars derived in this section are statistically consistent with those of type 1 quasars.

\subsection{Results}
\label{subsec:type2_results}

We derive the type 2 quasar luminosity function for three ranges in redshift: \za, {} \zb {} and \zc, using Eqs.~\ref{eq:phi} and \ref{eq:phierr}, summed over 420, 175 and 145 objects, respectively. Results with and without correction for the probability of detection of radio-selected objects (\S\ref{subsec:vvmax_alpha}) are shown in Fig.~\ref{fig:phitable_qso2_zbin}.

The flattening of the \zb {} and \zc {} LFs at low \oiiil {} luminosities is due to the decreased sensitivity of all the target algorithms to low-luminosity objects at high redshifts, mainly because of the algorithms' flux limits (see also \S\ref{subsec:type2_missed}). This incompleteness is partially mitigated by the radio-detection correction, but even this result is still strictly a lower limit. Therefore, for these redshift ranges, the most useful information from the derived LFs is provided by the highest luminosity bins. The trend is reversed for the \za {} LF, where the most useful information is provided by the low luminosity bins, since these represent many detected objects (see also \S\ref{subsec:type2_missed}).

While the high luminosity end of the LFs seem to suggest a tentative trend toward increasing number densities with increasing redshifts, we cannot make any conclusive statements because these are all lower limits. We return to the question of redshift evolution in the sample in \S\ref{subsec:test_vvmax}, but it is difficult to disentangle the effects of incompleteness of the sample with that of redshift evolution, and we make no attempt to do so.

\begin{figure}
\epsscale{0.9}
\plotone{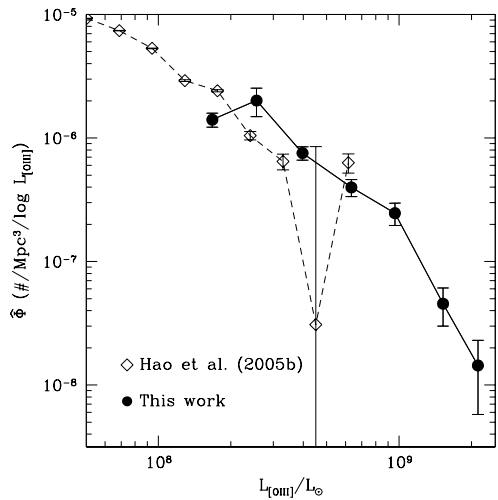}
\figcaption{\oiiil {} luminosity function of type 2 quasars (same as blue solid curve in Fig.~\ref{fig:phitable_qso2_zbin}) for \za, compared with that derived by \citet{hao05b}. We find good agreement between the two functions in the luminosity range in which they overlap. Here, the luminosities from \citet{hao05b} had been shifted up by 0.14 dex to account for the difference in their spectrophotometric flux calibration scale (see \S\ref{subsec:sdss_selec}).\label{fig:phitable_hao}}
\end{figure}

Our result is consistent with the type 2 AGN luminosity function in \citet{hao05b}, derived from objects with redshifts \za, selected from the SDSS Main Galaxy spectroscopic database. Figure~\ref{fig:phitable_hao} shows the high-luminosity end of this LF. We find good agreement with our derived type 2 quasar LF in the redshift and luminosity ranges in which the two samples overlap. In this Figure, the luminosities of \citet{hao05b} have been shifted up by 0.14 dex to account for the difference in their spectrophotometric flux calibration (see \S\ref{subsec:sdss_selec}). A more recent measurement of the type 1 AGN H$\alpha$ LF by \citet{gree07} is $\sim 40$ times lower at low luminosities than that derived by \citet{hao05b} because of their more stringent sample selection criteria, but the two LFs are consistent with each other at the high-luminosity end ($L_{\rmoiii}\sim 10^8L_{\odot}$), that is, at luminosities relevant to our comparison. 

\subsection{How Many Type 2 Quasars have we Missed?}
\label{subsec:type2_missed}

In principle, if we knew the distribution of optical SEDs and radio properties of type 2 quasars, we could calculate the probability with which each algorithm would select a type 2 quasar of a given [OIII]5007 luminosity at every redshift. We could then use this information to correct the luminosity function for objects that are not in our sample because they do not have the optical colors, apparent magnitudes, optical morphologies, or radio fluxes that are targeted by the SDSS algorithms. The optical spectrum of a type 2 quasar is the sum of the host galaxy spectrum, the narrow lines and scattered light from the AGN \citep{zaka05, zaka06}. None of these components can be neglected, the relative strengths of the components vary from object to object, and even the shape of the host galaxy continuum matters for selection. For example, the Balmer break of the host galaxy moving across the $r$ and $i$ filters at redshift $\sim 0.35$ produces a difference in colors sufficient to enable or disable selection by the Low-z QSO algorithm in several objects. This selection effect results in the weak feature in the \oiiil {} luminosity-redshift distribution seen in upper right-hand panel of Fig.~\ref{fig:lfcat}. 

With the above complications in mind, we do not attempt to model the spectra of type 2 quasars. Instead, we derive quantitative estimates of the contributory power of the various target algorithms to the selection of type 2 quasars, directly from the data. We do not use these results to apply corrections to the LF, in keeping with our `lower limit' approach; but they are nonetheless useful in guiding the interpretation of the derived luminosity functions. 

\begin{figure}
\epsscale{0.9}
\plotone{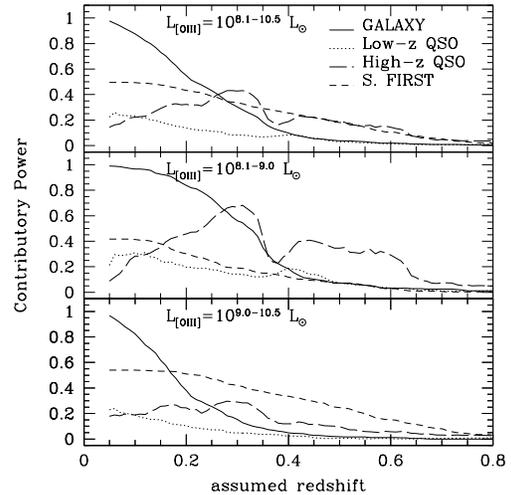}
\figcaption{Contributory power of the various SDSS target algorithms (as labeled)-- the fraction of objects in the type 2 quasar sample that would be selected by the algorithm taken alone. We place objects at different redshifts (in steps of $\Delta z = 0.01$) to determine this quantity as a function of assumed redshift. The three panels show the results for different ranges in \oiiil {} luminosity. All four algorithms are poor at selecting low \loiii {} objects at high redshifts, which is why the derived type 2 quasar LF shows evidence for incompleteness there.\label{fig:simflag_pselec_all}}
\end{figure}

We define the contributory power of a given target algorithm to be the fraction of objects in the sample (871 objects with \loiii $\ge 10^{8.1} L_\odot$) that we would select if we use this algorithm alone. We determine what this fraction would be if all objects were placed at a certain redshift (following \S\ref{subsec:type2_selec}). Figure~\ref{fig:simflag_pselec_all} shows the resulting function for assumed redshifts in the range of 0.1 to 0.8 (in steps of 0.01) and for three ranges in \oiiil {} luminosity. Not surprisingly, we find that the algorithms with deeper magnitude limits (Serendipity FIRST and High-z QSO) do better at selecting objects at high redshifts than do the shallow ones (Galaxy and Low-z QSO). The color-based selection criteria of the Low-z QSO and High-z QSO algorithms cause the peaks and dips in their curves. We find that all four algorithms are poor at selecting low \loiii {} objects at high redshifts, which is why the derived type 2 quasar LF shows evidence for incompleteness in this region.

\subsubsection{Contributory Power of Color-Based Target Algorithms}
An alternative approach to estimate the contributory power of the color-based target algorithms to the selection of type 2 quasars is to examine their overlap with the non-color based algorithms. Of the 258 Galaxy targets with redshifts $z<0.2$, 254 (98\%) are brighter than the magnitude limit of the Low-z QSO algorithm ($i_{\rm PSF}$ = 19.1) but only 12 of these (5\%) have colors that satisfy the Low-z QSO selection criteria. Between $z=0.2$ and $0.4$, there are 137 Galaxy targets; 113 of them (82\%) have $i_{\rm PSF} \le 19.1$, but only 1 object (0.9\%) was selected by the Low-z QSO algorithm. The small amount of overlap between the Low-z QSO and Galaxy targets indicates that the Low-z QSO algorithm is missing many type 2 quasars because it is only sensitive to a limited region of color space. There is no overlap at all between the High-z QSO and Galaxy algorithms. The colors of many type 2 quasars are dominated by the colors of their host galaxies, especially for lower [OIII] luminosities, and neither Low-z QSO nor High-z QSO algorithms target such objects.

Out of the 275 objects selected by the Serendipity FIRST target algorithm, 74 have $i_{\rm PSF} \le 19.1$ and 15 of these (20\%) are targeted by the Low-z QSO algorithm; 245 have $i_{\rm PSF} \le 20.2$ and only 12 of these (5\%) are targeted by the High-z QSO algorithm. These comparisons suggest that over the full redshift range, the Low-z QSO and High-z QSO algorithms alone select only 20\% or less of type 2 quasars. 

\subsubsection{Low-redshift Objects and the Galaxy Target Algorithm}
\begin{figure}
\epsscale{0.9}
\plotone{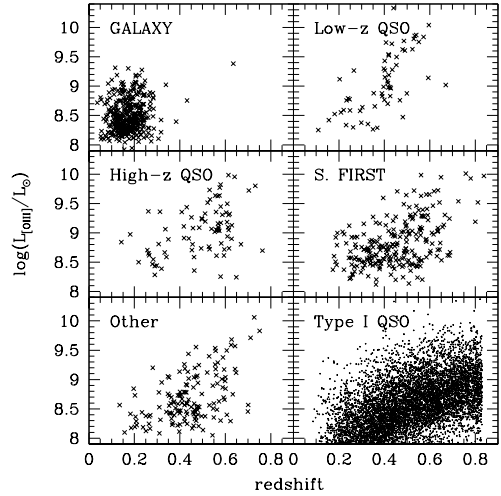}
\figcaption{Distribution of \oiiil{} luminosities and redshifts of type 2 quasars targeted by the Galaxy (366 objects), Low-z QSO (76), High-z QSO (89), Serendipity FIRST (276), and other (147) target algorithms. Some objects are targeted by multiple algorithms and thus appear in multiple panels in this Figure. Also shown is the distribution for type 1 quasars (8003 objects; see \S\ref{subsec:type1_samp}). \label{fig:lfcat}}
\end{figure}

In this section, we discuss the interpretation of our LF calculation in the lowest redshift range, \za. We see from Fig.~\ref{fig:lfcat} that there are no objects in our sample with \loiii $>10^{9.5} L_\odot$ found at these redshifts, and only 1 out of 17 objects with \loiii $>10^{9.8}L_\odot$ is at $z \le 0.5$. At face value, the scarcity of detected objects may be interpreted as a significant drop in the ratio of number densities of type 2 quasars at these luminosities. However, we argue here that this may instead be due to a combination of the selection effects and the small volume covered by the lowest redshift bin. 

Most type 2 quasars in the redshift range $z<0.3$ were selected by the Galaxy algorithm. The efficiency of the Galaxy algorithm begins to drop at redshifts around $z=0.2$ due to its shallow magnitude limit (Figs.~\ref{fig:lfcat} and \ref{fig:simflag_pselec_all}). Quantitatively, an object with \loiii$=10^{9.5} L_\odot$ at $z=0.3$ with a rest-frame equivalent width of 500\AA\ (corresponding to the median of the \loiii/EW$_{\rm [OIII]}$ distribution for our sample; Paper I) would have an $r$ band AB magnitude of around 17.8 mag, and would therefore not be selected by the Galaxy algorithm (which has a limiting magnitude of 17.77 mag). As we discussed in the previous section, these objects may be missed by the color-based target algorithms because they do not lie in the specific regions of color-color space these algorithms were designed to target. The radio-based Serendipity FIRST algorithm would have a chance of identifying high-luminosity, low-redshift objects that happen to be radio-bright, but even the luminous objects at low redshift may fall below the radio flux limits of the FIRST survey. More importantly, only $\sim$ 40\% of Serendipity FIRST targets have observed spectra (\S \ref{subsec:vvmax_alpha}). Therefore, the fact that we have no type 2 quasars with \za {} and \loiii $>10^{9.5} L_{\odot}$ may be entirely due to the incompleteness of SDSS spectroscopic target selection. 

\section{Type 1 Quasar Luminosity Function}
\label{sec:type1_lf}

\begin{figure}
\epsscale{0.9}
\plotone{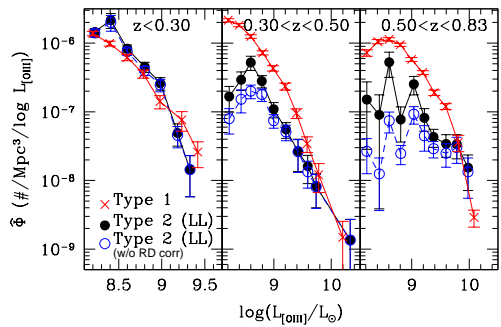}
\figcaption{Comparison of the \oiiil {} luminosity functions of type 1 and type 2 quasars, for three ranges in redshift. The type 2 quasar luminosity function \textbf{(lower limits)} with and without the correction for the probability of radio detection (see \S\ref{subsec:vvmax_alpha}) are shown (black solid and blue open circles, respectively). The type 2 quasar sample suffers from incompleteness, especially at low \oiiil {} luminosities and high redshifts. Nevertheless, we find that the derived space densities of type 1 and type 2 quasars are comparable for the redshift range \za {} and for the highest luminosities for the higher redshifts $0.3<z<0.83$. \label{fig:phitable_qso12_union_zbin}}
\end{figure}

In order to put our type 2 quasar LF into context, we need to compare it with the type 1 quasar LF at similar redshifts and luminosities. In this section, we derive the \invvmax {} \oiiil {} LF from a complete sample of type 1 quasars using the formalism we have applied to the type 2 quasar sample (\S\ref{subsec:type2_invvmax}). We discuss the sample selection in \S\ref{subsec:type1_samp}, measurement of \oiiil {} luminosities in \S\ref{subsec:type1_o3}, the calculation of the LF and results in \S\ref{subsec:type1_invvmax}, and comparison with 
previous work in \S\ref{subsec:type1_rich}. The type 1 LF for the three redshift ranges is shown in Fig.~\ref{fig:phitable_qso12_union_zbin}. In the following section, \S\ref{sec:type12}, we combine this result with the derived type 2 quasar LF to calculate the ratio of space densities of the two populations. 
 
\subsection{Type 1 Quasar Sample}
\label{subsec:type1_samp}

We select a complete sample of 8,003 type 1 quasars from the SDSS DR5 Quasar Catalog \citep{schn07}. The catalog consists of 77,429 spectroscopically-confirmed quasars in the redshift range $0.08$--$5.41$, with absolute magnitude in the $i$ band, $M_i< -22$ mag and full width at half-maximum (FWHM) of lines from the broad-line region greater than 1000 km s$^{-1}$. From this catalog, 31,999 objects are targeted with the final version of the Low-z QSO algorithm \citep{rich02}. Of these, 8,003 objects satisfy the redshift cut $z<0.83$ and are included in our sample. The effective area of our sample is 4,041 deg$^2$, calculated using the same procedure as in \S\ref{subsec:type2_invvmax}. 

The redshift and \oiiil {} luminosities of these objects are shown in the bottom-right panel of Fig.~\ref{fig:lfcat}. There is no explicit lower cut in redshift, but there are very few objects with redshifts $z<0.15$ because of the strong redshift evolution in the type 1 quasar population and the bright limit of the Low-z QSO target algorithm.

\subsection{Measurement of \oiiil {} Luminosities}
\label{subsec:type1_o3}

In the calculation of the type 1 quasar LF, we use \oiiil {} luminosities measured from Gaussian fits, after careful subtraction of \feii {} emission. The average contribution of \feii {} emission to the total flux over the wavelength range $5007 \AA \pm 1.7 \times$ FWHM is $\sim$ 9\%. We exclude the region containing \hb {} and \oiii 4959,5007 and find the best fit to the spectrum in the form of a power law plus the \feii {} template of \citet{boro92}. We then fit a set of four Gaussians to  \oiii 4959, \oiii 5007, and narrow and broad components of H$\beta$. 5,297 quasars (67\% of the sample) have \loiii $> 10^{8.3} L_\odot$, and 221 quasars (2.8\% of the sample) have \loiii $> 10^{9.5} L_\odot$. 

We also obtain a non-parametric line luminosity by integrating the detected flux density and subtracting the contribution from the continuum, \feii, \hb, and \oiii 4959. For most objects, the Gaussian and integrated line fluxes are in good agreement. Outliers from the integrated vs. $\!$Gaussian locus were visually inspected, and 187 spectra (2.3\% of the sample) for which the fit failed (due to low signal-to-noise ratio, bad pixels, etc.) were discarded. 

\subsection{\invvmax {} Luminosity Function}
\label{subsec:type1_invvmax}

Objects in the type 1 quasar sample were all selected by the Low-z QSO target algorithm, described in \S\ref{subsec:vvmax_qso}. In the case of type 2 quasars, we considered both the magnitude limits and color-based selection criteria. In contrast, type 1 quasar colors do not evolve strongly with redshift for $z < 2.2$ \citep{rich01,rich02}, so we do not expect the color-based selection criteria to be important in this case. For each object, we determine the available comoving volume \vmax {} based only on the faint magnitude limit of the survey, $i_{\rm PSF} = 19.1$, since the volume corresponding to the bright limit of the survey is negligibly small. 

As in \S\ref{subsec:type2_selec}, we calculate scaled and k-corrected PSF magnitudes as a function of assumed redshift from the observed spectrum of each object. We determine $z_{\rm max}$, the redshift at which the PSF magnitude reaches the limit of the survey. The available volume \vmax {} is then given by
\begin{equation} 
V_{\rm max} = \frac{\Omega}{4\pi} \int_{z_1}^{\min(z_{\rm max},z_2)}  \frac{dV_{\rm c}}{dz}(z){}  dz,\label{eq:vmax2}
\end{equation}
where $\Omega = 4041 \mbox{ deg}^2$  is the effective survey area for this sample, $z_1$ and $z_2$ are the edges of the redshift range for which we are calculating the LF and $(dV_{\rm c}/dz) \, dz$ is the comoving volume element in the redshift interval $dz$.

\begin{figure}
\epsscale{0.9}
\plotone{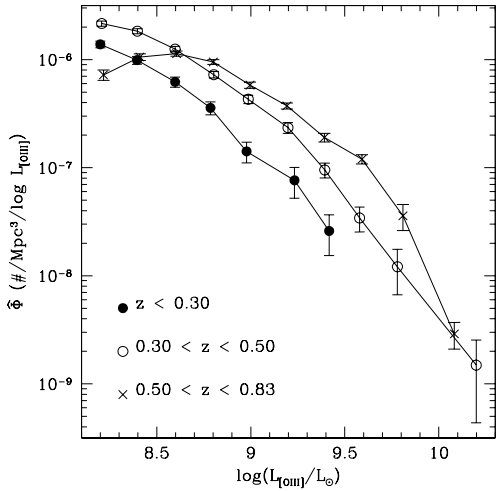}
\figcaption{\oiiil {} luminosity function of type 1 quasars for three ranges in redshift. The maximum volume is calculated using the magnitude limit of the Low-z QSO target algorithm. The turn-over of the \zc {} LF at low luminosities is an artifact due to the difficulty of measuring weak \oiiil {} lines at these high redshifts. \label{fig:phitable_qso1_zbin}}
\end{figure}

Figure \ref{fig:phitable_qso1_zbin} shows the \invvmax {} \oiiil {} luminosity function for type 1 quasars (calculated using Eqs.~\ref{eq:phi} and \ref{eq:phierr}), for the same redshift ranges as for type 2 quasars. Our results reflect the positive redshift evolution of the type 1 quasar population \citep{rich06}. The turn-over of the \zc {} LF at low luminosities is an artifact due to the difficulty of measuring weak \oiiil {} lines at these redshifts, since the line falls on the red end of the observed spectrum where the signal tends to be noisier. 
\subsection{Comparison with the Broadband Type 1 Quasar LF}
\label{subsec:type1_rich}

In this section, we test whether our measured \oiiil {} type 1 quasar LF is consistent with the observed broadband LF, based on type 1 quasars from the SDSS DR3 quasar catalog, with redshifts $0.30<z<0.68$ (Table~6 of \citealt{rich06}). The measure of luminosity used there is the continuum luminosity around rest-frame wavelength 2500\AA, corresponding to the SDSS $i$ band at $z=2$ (e.g., \citealt{blan03}), expressed as a broadband absolute magnitude $\Mi$. This test serves two purposes. First, it is a sanity check for our LF calculation. Second, it probes the correlation between \oiiil {} line luminosity and continuum luminosity in type 1 AGN. This is important for justifying our assumption that \oiiil {} luminosity traces bolometric luminosity sufficiently well for our purposes (see \S\ref{subsec:disc_o3}). As expected, we find a strong correlation between $\Mi$ and \oiiil {} luminosity in our type 1 quasar sample (Fig.~\ref{fig:o3_Mi}). The mean relation is
\begin{equation} \label{eq:o3_Mi}
\log \left(\frac{L_{\rm [OIII]}}{L_\odot}\right) = -0.38 \Mi -0.62,
\end{equation}
consistent with a linear relation, with a slope of 0.95 in the $\log L_{2500}$--$\log L_{\rm [OIII]}$ plane. The scatter in \loiii {} at fixed $\Mi$ is consistent with a log-normal distribution of width 0.36 dex. 

\begin{figure}
\epsscale{0.9}
\plotone{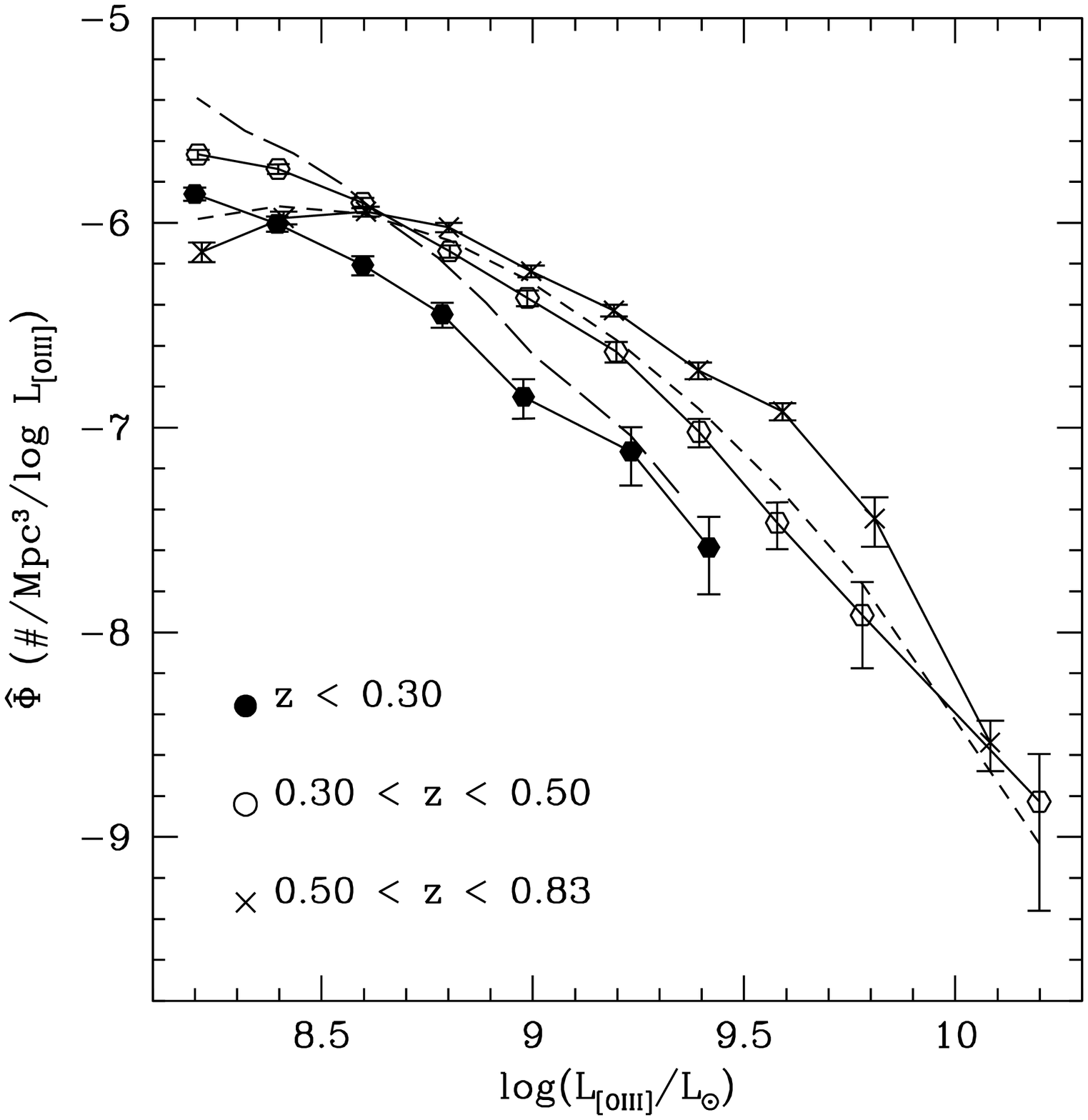}
\figcaption{Comparison of the derived \oiiil {} luminosity function of type 1 quasars with previous work. We convolve the broadband LF derived by \citet{rich06} for $0.30<z<0.68$ with the mean \loiii-$\Mi$ relation Eq.~\ref{eq:o3_Mi} with a log-normal scatter of 0.36 dex. The converted \oiiil {} luminosity function (short-dashed curve) is in good agreement with the derived LF in the redshift range \zb. The shape of the converted LF is sensitive to the assumed scatter in the \loiii-$\Mi$ relation. Also shown is the converted \oiiil {} LF derived under the assumption of zero scatter (long-dashed curve), which is inconsistent with the derived LF. \label{fig:phitable_o3richards}}
\end{figure}

To convert the broadband LF, $\Phi(\Mi)$, into an \oiiil {} LF, $\Phi_{\rm conv}$(\loiii), we convolve the broadband LF with the mean \loiii-$\Mi$ relation Eq.~\ref{eq:o3_Mi}, with a log-normal scatter of width $\sigma = 0.36$ dex, i.e.,
\begin{equation}
\Phi_{\rm conv}({\cal L}) = \int \Phi(M) \exp\left[-\frac{({\cal L} - \overline{\cal L}(M))^2}{2\sigma^2}\right] d\,M,
\end{equation}
where ${\cal L} = \log(L_{\rm [OIII]}/L_\odot)$ and $M = \Mi$. Figure~\ref{fig:phitable_o3richards} shows that there is a good agreement between the converted LF (short-dashed curve), and our measured \oiiil {} LF for \zb {} (solid line, open circles). The turn-over at the low \loiii {} end of the converted LF is due to the absolute magnitude cut-off, since there are no objects fainter than $\Mi \sim -23$ mag to scatter into these \loiii{} bins. 

The shape of the converted LF is sensitive to the assumed slope and scatter in the \loiii--$\Mi$ relation. In particular, if zero scatter in the \loiii-$\Mi$ relation is assumed, the resulting converted \oiiil {} LF is inconsistent with our derived LF (long-dashed curve). With scatter properly taken into account, we find that the two independent derivations of the LF match.  This suggests that we can successfully convert an \oiiil {} LF to a broadband LF (and vice versa). 

\section{Comparison of Type 1 and Type 2 Quasar Luminosity Functions}
\label{sec:type12}

In this section, we combine the results of the previous sections to compare the type 1 and type 2 quasar populations. Most importantly, we can now use the derived \oiiil {} type 1 and type 2 quasar luminosity functions to constrain the fraction of type 2 quasars as a function of \oiiil {} luminosity. We discuss the calculation, results and their interpretation in \S\ref{subsec:type2_frac}. We compare the two populations in terms of their redshift evolution in \S\ref{subsec:test_vvmax}, and their radio properties in \S\ref{subsec:test_radio}.

\subsection{Type 2 Quasar Fraction}
\label{subsec:type2_frac}

Figure~\ref{fig:phitable_qso12_union_zbin} shows that the \oiiil {} type 1 and type 2 quasar luminosity functions are comparable in the regimes where we expect our type 2 quasar LF to be least affected by selection bias, i.e., the low-luminosity regime ($<10^9 L_\odot$) for the redshift range \za {} and the high-luminosity regime for redshift ranges \zb {} and \zc. From these two functions, we directly calculate the type 2 quasar fraction, i.e., the ratio of type 2 to total (type 1 $+$ type 2) quasar number densities. We determine the number density of quasars in a given luminosity range by integrating the LF over that range. For this calculation, we use our best lower bound to the type 2 quasar LF (which includes a correction for the probability of radio detection, \S\ref{subsec:vvmax_alpha}). 

\begin{figure}
\plotone{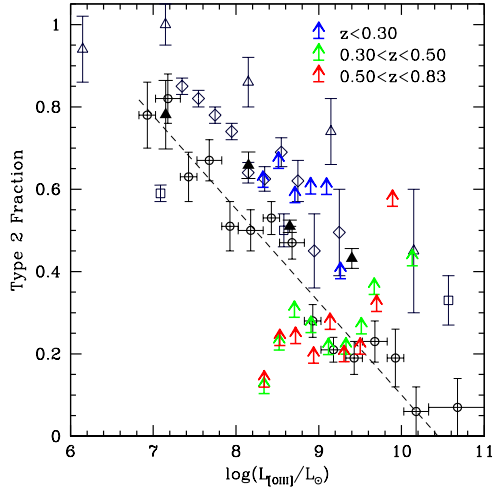} 
\figcaption{Lower limits to the type 2 quasar fraction: the ratio of type 2-to
total (type 1 + type 2) quasar number densities for three ranges in
redshift: $z<0.30$ (blue), $0.30<z<0.50$ (green), and $0.50<z<0.83$
(red). Number densities are estimated by integrating the LF over bins
in \oiiil\, luminosity. We have used our best lower bound to the type
2 quasar LF (which includes a correction for the probability of radio
detection; see Section 3.2.4 in the main paper). Obscured quasar
fractions derived from X-ray surveys are from Ueda et~al. (2003; grey
open squares) and Hasinger (2008; black open circles and dashed line); 
those derived from IR data are Treister et~al. (2008; black filled
triangles). Also shown are results from radio data (Grimes
et~al. 2004; grey open triangles) and optical data (Simpson
et~al. 2005; grey open diamonds).\label{fig:type2_frac}} 
\end{figure}

Figure~\ref{fig:type2_frac} shows the calculated type 2 quasar fraction as a function of \oiiil {} luminosity. For the \za {} bin, we find that the type 2 quasar fraction is about 60\% for \loiii $=10^{8.3}$ to $10^9 L_\odot$. For \za {} and \zb, our strongest limits come from the highest luminosity bin, which constrain the type 2 quasar fraction to be at least 40\% and 60\% at \loiii {} $\sim 5 \times 10^{9} L_\odot$, respectively. 

Our derived type 2 quasar fraction is a lower limit for the following reasons: (i) our derived type 2 LF is a lower limit (see beginning of \S\ref{sec:type2_lf}); (ii) the combination of selection effects and redshift evolution artificially lowers the type 2 quasar fraction that we derive (as we describe in \S\ref{subsec:test_vvmax}); (iii) there are indications that the \oiiil {} line is slightly more extincted in type 2 quasars than in type 1 quasars (\S\ref{subsec:test_radio} and \S\ref{subsec:reddening}); and (iv) we have assumed that the \oiiil {} luminosity is independent of the obscured fraction and serves as a tracer of the bolometric luminosity. In practice, it is likely that the \oiiil {} luminosity is higher for objects with a larger opening angle, since in this case more material is illuminated by the central source. As demonstrated by \citet{krol99}, this dependence leads to a bias favoring unobscured objects in any [OIII]-flux selected sample.

There has been a substantial amount of work to determine the obscured quasar fraction at different luminosity and redshift regimes. Some of these results are shown together with ours in Fig.~\ref{fig:type2_frac}. Nearest to our approach are previous determinations from emission-line luminosity functions of low-luminosity (type 1 and type 2) AGN in the SDSS, at redshifts $z \sim 0.1$. \citet{hao05b} found that type 2 AGN make up about 60\% of the AGN population at $\mbox{\loiii} \sim 10^6 L_\odot$, and about 30\% at $\sim 3 \times 10^{7} L_\odot$. \citet{simp05} also clearly finds the decreasing trend in type 2 fraction with luminosity, but found substantially higher type 2 fractions (open diamonds).

Determinations from hard X-ray (2-10 keV) selected samples, at
redshifts $z \sim 3$, suggest that the fraction of obscured quasars is
large at low luminosities, and then decreases at higher luminosities
\citep{ueda03, szok04, barg05, mark05, trei06, beck06, sazo07}.
Figure~\ref{fig:type2_frac} shows results from Ueda et. al. (2003;
open squares) and Hasinger (2008; open circles and dashed line).
We have converted the hard (2-10 keV) X-ray luminosities to \oiiil {}
luminosities by shifting by 1.59 dex (toward lower luminosities)
following \citet{heck05}. X-ray surveys may be missing Compton-thick
objects which might constitute a significant fraction of all AGN
\citep{heck05,poll06,mart07}, so these values should also be treated
as lower limits to the true obscured quasar fraction. Also shown are
results from radia data (Grimes et al. 2004; open triangles).
 
From samples selected from infrared observations, \citet{mart06} estimated the fraction of obscured quasars at $z\sim 2$ to be $\sim$70\%, with a large uncertainty because of small sample size. \citet{poll07} estimated the fraction of obscured quasars to be $60-65\%$ at bolometric luminosities $10^{46-47} \mbox{erg s}^{-1}$ (\loiii $\sim 10^9 L_\odot$), using a sample of type 2 quasars with Si absorption. Again, these values should be taken as lower limits because, e.g., the former sample excludes radio-weak objects, while the latter excludes sources with featureless IR spectra \citep{stur06}. 

Recently, \citet{trei08} determined the type 2 quasar fraction by
calculating the fraction of the total light emitted by obscuring
material in the IR in type 1 AGN (filled triangles).
Shown are the results assuming the parameter $f_{12} = 0.06$, the fraction of the total dust-reprocessed luminosity falling within the MIPS band (11-13 $\mu$m in the rest frame at $z \approx 1$), and with model-dependent error bars. We have converted their quoted bolometric luminosities to \oiiil {} luminosities using the approximate conversion in \S\ref{subsec:disc_o3}. 

Overall, our lower limits on the type 2 quasar fraction are consistent with previous determinations. In practice, however, we suspect that the true type 2 quasar fraction is significantly higher than our lower limits, for the reasons discussed above and in \S\ref{sec:discussion}. This simply underscores the need to treat most determinations of the obscured quasar fraction as lower limits, since a substantial part of the population may be missed by wavelength- and method-specific selection criteria.

\subsection{Comparison of Redshift Evolution}
\label{subsec:test_vvmax}

In this section, we apply the \vvmax {} test \citep{schm68} to probe redshift evolution within the type 1 and type 2 samples. For a uniformly-selected, non-evolving population of objects, \vvmax {} is expected to be distributed uniformly between 0 and 1 and to have a mean value of 0.5. For an object with redshift within the range $(z_1,z_2)$, \vvmax {} is given by
\beq
\frac{V}{ V_{\rm max}}= \frac{ \int^{z_{\rm obs}}_{z_1} P_{\rm s}(z)  \frac{dV_{\rm c}}{dz}(z){}  dz }{ \int_{z_1}^{z_2} P_{\rm s}(z)  \frac{dV_{\rm c}}{dz}(z){}  dz}
\eeq 
where $z_{\rm obs}$ is the observed redshift of the object and $(dV_{\rm c}/dz){} dz$ is the comoving volume element in the redshift interval $dz$. If there is positive redshift evolution within the redshift range, the mean values of \vvmax {} over the sample are greater than 0.5. Incompleteness in the sample can either mask or enhance this effect, by weighting the sample toward the lower or higher end of the redshift range.

In Table~\ref{tab:vobsvmax}, we list the mean values of \vvmax {} calculated from the type 1 and type 2 quasar samples for the three redshift ranges we used previously. Our type 1 quasar sample is complete, and the \mvvmax {} values greater than 0.5 indicate the positive redshift evolution within the sample \citep{rich06}. On the other hand, our type 2 quasar sample is incomplete, so the deviation of \mvvmax {} from 0.5 is due to a combination of redshift evolution and selection effects (see \S\ref{subsec:type2_missed}), which we do not attempt to disentangle. For two out of the three redshift ranges, \mvvmax {} values are lower for the type 2 quasar sample than for the corresponding type 1 quasar sample.

Since our derived LFs are calculated for broad ranges in redshift, they are a weighted average of the true LFs over the redshifts within this range. The above results suggest that the type 1 quasar LFs are weighted toward the higher end of the redshift range, and therefore toward higher number densities, relative to the type 2 quasar LFs. Therefore, the combination of selection effects and redshift evolution of quasar number density artificially lowers the type 2 to type 1 ratio that we derive from these LFs.

\subsection{Comparison of Radio Properties}
\label{subsec:test_radio}

In this section, we test whether the radio properties of type 2 quasars are statistically consistent with those of type 1 quasars. We test this hypothesis to the same limited extent that we have employed in \S\ref{subsec:vvmax_alpha} -- that is, we test whether non-radio-selected type 2 and type 1 quasars have the same detection rate with FIRST within a matching radius of 2\arcsec. We use a variation of the statistic defined in Eq.~\ref{eq_u} for comparing two populations of objects:
\begin{equation} \label{eq_u2}
u=\frac{(f_1-f_2)^2}{\sqrt{f_1(1-f_1)f_2(1-f_2)}}\frac{N_1 N_2}{N_1+N_2}.
\end{equation}
Here $N_1$ and $N_2$ are the number of type 1 and type 2 quasars, respectively, in a given bin in the $z$-\loiii {} plane and $f_1$ and $f_2$ are the fractions of objects in this bin that are detected by FIRST. The distribution of the statistic described by Eq.~\ref{eq_u2} (inspired by two-sample statistics from \citealt{peac83}) is not formally independent of $f_i$ and $N_i$ in the same sense that the distribution of Eq.~\ref{eq_u} is independent of $N$ and $p$ (as shown in the Appendix). Using Monte-Carlo simulations we found that the distribution of the value in Eq.~\ref{eq_u} is very close to that of $\chi^2$ with one degree of freedom, as long as $f_1$ and $f_2$ are not too close to 0 or 1 and as long as the number of objects in each bin is sufficiently large ($N_i\ga 10$). 

We find that the radio-detection rates of the two populations are statistically indistinguishable for all except the low-redshift, low-luminosity objects. Type 2 quasars with $z<0.25$, $8.0<\log(L_{\rm [OIII]}/L_{\odot})<8.5$ are significantly more likely to be detected in FIRST than type 1 quasars. One possibility is that the \oiiil {} emission line is systematically more extincted by dust in type 2 quasars than in type 1 quasars. Correcting for 0.1 dex of this putative extinction removes the detection rate difference everywhere in the $z$-\loiii {} plane. 


\section{Discussion}
\label{sec:discussion}

We have presented our main results in terms of \oiiil {} luminosity, a proxy for AGN activity that is directly measurable from the observed spectra. In this section, we discuss two issues that are important for the interpretation of these results: the relation between \oiiil {} and bolometric luminosity and the amount of extinction of the \oiiil {} line (\S\ref{subsec:disc_o3} and \ref{subsec:reddening}, respectively). We conclude that despite these effects, our approach to calculate lower limits on the type 2 quasar luminosity function and the type 2 quasar fraction remains valid. In \S\ref{subsec:soltan}, we discuss implications of our results for estimates of the accretion efficiency of supermassive black holes using measurements of AGN space densities. 


\subsection{\oiiil {} Luminosity as Tracer of Bolometric Luminosity}
\label{subsec:disc_o3}

The \oiiil {} emission line luminosity is arguably the best available proxy for AGN activity for optically-selected obscured quasars. This line is emitted by the narrow-line region, which extends outside the obscuring material thought to surround the broad-line region of type 2 AGN \citep{anto93,urry95}. Observational support for this assumption comes from the similarity of IR/[OIII] ratios in type 1 and type 2 AGN found by \citet{mulc94}. Moreover, the \oiiil {} line has the advantage of being strong and easily detected in most AGN, and \citet{simp98} and \citet{kauf03} have shown that it is a good tracer of AGN activity and is not severely contaminated by star formation.

\begin{figure}
\epsscale{0.9}
\plotone{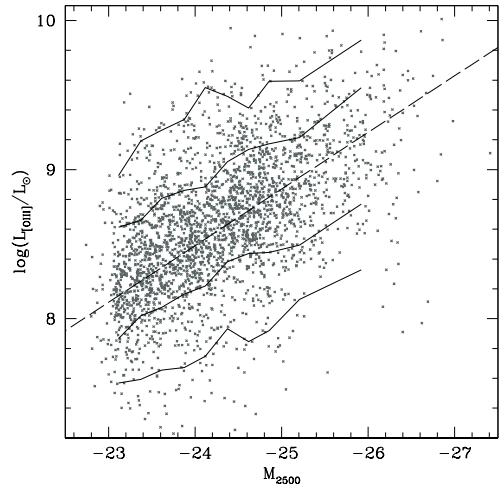}
\figcaption{Correlation between broadband absolute magnitude $\Mi$ and \oiiil {} luminosity for type 1 quasars in our sample. $\Mi$ measures the continuum luminosity at around 2500~\AA, corresponding to the SDSS $i$ band filter at $z=2$. We find the mean relation to be $\log (L{\rm [OIII]}/ L_\odot) = -0.38 \Mi -0.62$ (dashed line), i.e., the relation is close to linear. The scatter in \loiii {} at fixed continuum luminosity is consistent with a log-normal scatter with a width of 0.36 dex; 1-sigma and 2-sigma contours are shown here (solid curves). \label{fig:o3_Mi}}
\end{figure}

For our type 1 quasar sample, we found that \oiiil {} luminosity correlates strongly with broadband luminosity at around 2500 \AA {} (Fig.~\ref{fig:o3_Mi}), as well as with the monochromatic continuum luminosity measured at rest-frame 5100\AA {} (linear correlation coefficient = 0.53). More importantly, we have demonstrated in \S\ref{subsec:type1_rich} that given the observed mean and scatter of the correlation between \oiiil {} and continuum luminosity allows to quantitatively convert the type 1 quasar broadband LF to a \oiiil {} LF that is consistent with our independent measurement.

Of course, \oiiil {} luminosity is not a perfect tracer of bolometric luminosity. Indeed, there is substantial scatter between \oiiil {} and continuum luminosity for type 1 quasars (see Fig.~\ref{fig:o3_Mi}; \citealt{netz06} and references therein). It is plausible that this scatter reflects a real, physical difference in covering fractions of the narrow-line region. As mentioned in \S\ref{subsec:type2_frac}, this would imply that our [OIII]-flux selected sample is biased toward type 1 sources and would artificially lower our derived type 2 quasar fraction. This is in line with our approach to calculate lower limits on this quantity.

Rough estimates of bolometric luminosities can be found by, first, using the $M_{2500}-L_{\rmoiii}$ relation (Eq.~\ref{eq:o3_Mi}), and then the well-measured average type 1 quasar spectral energy distributions. For example, an \oiiil {} luminosity of $3\times 10^{9}L_{\odot}$ corresponds to an intrinsic UV luminosity of $M_{2500}\approx -26.6$ mag. Next, we use the average quasar SED from \citet{vand01} to calculate the corresponding luminosity in the $B$ band. Finally, we use bolometric corrections from \citet{marc04}, derived from a template AGN spectrum and with model 1-sigma uncertainties of $\sim$ 0.05 dex. We estimate that the corresponding bolometric luminosity is $\sim 1\times 10^{47}$ erg s$^{-1}$ or $3\times 10^{13} L_\odot$. 


\subsection{Reddening and Extinction}
\label{subsec:reddening}

The \oiiil {} line, which we use as a proxy for AGN activity, is emitted by an extended narrow-line region. It is expected to be affected by extinction due to interstellar dust, located either within the narrow-line region itself or in the intervening interstellar matter of the host galaxy. In this section, we estimate the magnitude of this extinction for type 2 quasars and discuss its effects on our measurements of the luminosity functions and the type 2 quasar fraction. 

We attempt to estimate narrow-line region extinction for the luminous subsample of type 2 quasars (with \loiii$>10^9 L_\odot$) by determining their Balmer line ratios. We measure H$\alpha/$H$\beta$ for 44 objects (with $z<0.4$, for which the H$\alpha$ line is observed) and H$\beta$/H$\gamma$ for 203 objects (for which the H$\gamma$ line has sufficiently high S/N). We use fluxes measured using the non-parametric line fitting procedure described near the end of \S\ref{subsec:sdss_selec} and illustrated in Fig.~\ref{pic_halpha}. 

\begin{figure}
\plotone{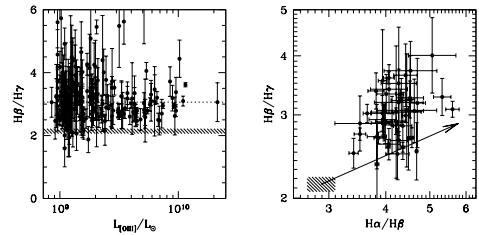}  
\figcaption{Balmer line ratios of luminous type 2 quasars, for objects in which the relevant emission lines can be measured. Left panel: H$\beta$/H$\gamma$ vs. \oiiil {} luminosity (203 objects). The median H$\beta$/H$\gamma$ is 3.06 (dotted line), which corresponds to an \oiiil {} extinction of 2.6 mag (for a Milky Way extinction curve and a foreground obscuring screen). Right panel: H$\alpha/$H$\beta$ vs. H$\beta$/H$\gamma$ (44 objects). The arrow shows the extinction vector for the Milky way extinction curve, whose length corresponds to an \oiiil {} extinction of 2.2 mag. In both panels, the hatched area shows case A--B ratios at temperatures 5,000--20,000 K, and errors are estimated by comparing non-parametric and Gaussian flux measurements.\label{fig_balmerdecr}}
\end{figure}

Our results are shown in Fig.~\ref{fig_balmerdecr}. The left panel shows H$\beta$/H$\gamma$ vs.\,\oiiil {} luminosity. The median value of H$\beta$/H$\gamma$ is 3.06 (dotted line), which corresponds to an extinction of \oiiil {} of 2.6 mag for a Milky way extinction curve with $R=3.1$ and a foreground obscuring screen. Error bars are based on the difference between the flux measured from our non-parametric fitting procedure and that derived from a Gaussian fitting procedure or from integration over the emission line (after subtraction of the continuum). The right panel shows H$\alpha/$H$\beta$ vs. H$\beta$/H$\gamma$ for the subset of objects with both measurements. The arrow shows the extinction vector for the Milky Way extinction curve, whose length corresponds to an \oiiil {} extinction of 2.2 mag. Other commonly used reddening laws (e.g., the Small Magellanic Cloud extinction curve) yield very similar extinction vectors. The data are clearly inconsistent with standard reddening laws. For example, the median H$\alpha/$H$\beta=4.25$ corresponds to an extinction of 1.2 mag, which is $\sim$ 1.4 mag lower than that derived from the median H$\beta$/H$\gamma$. There is no statistical evidence for a correlation between \hahb {} and \hbhg {} (correlation coefficient $=0.3$). Furthermore, the data are inconsistent with a linear regression from the case A--B values, and therefore cannot be described by a dust screen model, no matter what the extinction law is. 

Early studies of Balmer line intensities of quasars have indicated that these cannot be explained by standard recombination theory and a standard dust reddening law \citep{ande70,adam75,bald75,oste76,oste77}. Other processes that may play an important role include collisional excitation from higher energy levels, self-absorption of the Balmer lines, continuum optical depth, scattering and fluorescence \citep{capr64a,capr64b,netz75,krol78}. Given the complications, we do not apply dust extinction corrections to our  \oiiil {} luminosity measurements. 

Our use of extinction-uncorrected line luminosities means that in our calculation of the luminosity function, objects with \oiiil {} lines affected by extinction have been assigned to fainter luminosity bins. Hence, the LF has been shifted toward fainter luminosities, or equivalently, toward lower space densities, relative to an  extinction-corrected LF. Moreover, if line extinction is stronger in type 2 than in type 1 sources, then the type 2 quasar fraction would also be underestimated. Several independent measurements have suggested that this is indeed the case \citep[][and references there in]{netz06}. Reddening estimates from Balmer decrements suggest that the narrow-line regions of type 2 sources are more heavily reddened than those of type 1 sources \citep{dezo85,daha88}. Recent IR observations strongly indicate that the narrow-line regions of type 2 quasars are significantly more obscured than those of type 1 quasars, by up to a factor of 10 \citep{haas05}. \citet{netz06} suggest that \oiiil {} luminosity is a factor of $\sim$ 2 larger in type 1 than in type 2 quasars for the same 2--10 keV X-ray luminosity, as would be the case if type 2 sources are more reddened than type 1 sources. Finally, such a trend is also hinted at by our analysis of radio properties of type 1 and type 2 sources (\S\ref{subsec:test_radio}). Therefore, the use of luminosities uncorrected for extinction is consistent with our approach to calculate lower limits on both the type 2 quasar LF and the type 2 quasar fraction. 

\subsection{Implications for the Black Hole Mass Function and Accretion Efficiency}
\label{subsec:soltan}

If black holes grow primarily by accretion, the total luminosity emitted 
by accretion processes over the lifetime of the Universe is directly 
related to the accumulated mass of local supermassive black holes 
\citep{solt82,yu02}. Recent measurements of 
both the mass density of supermassive black holes and the total luminosity 
emitted by matter accreted onto them can be reconciled if the radiative 
efficiency (the fraction of the accreted mass that is converted directly 
into observed quasar radiation) is quite high, $\varepsilon\ga 0.1$ 
\citep{yu02, marc04, barg05}. In this section we discuss qualitatively how 
these calculations are affected by the existence of a large population of 
obscured quasars.

What matters for quantifying the growth of the black hole is its total 
energetic output (`bolometric luminosity' $L_{bol}$), since that is the 
value that can be related to the accreted mass through radiative 
efficiency. Some of the optical, UV and X-ray radiation is intercepted by 
the obscuring material and then re-emitted in the IR more or less 
isotropically, but no new energy is generated. The fraction of energy 
reprocessed this way roughly equals $f(L_{bol}) L_{bol}$, where 
$f(L_{bol})$ is the fraction of obscured AGN in the population and $4\pi 
f(L_{bol})$ is the solid angle covered by obscuring material as seen from 
the central engine. Since the IR emission is directly related to the 
obscuration fraction, it is possible to derive the obscured fraction from 
the IR-to-optical ratio in unobscured quasars \citep{trei08}.

Some of the accretion efficiency calculations are based on the optical LF 
of type 1 quasars (e.g., \citealt{yu02}). In this case, the luminosity of 
each individual quasar is overestimated by including the IR emission in 
the bolometric correction, but the total number of objects is 
underestimated. The net result is that the contribution of quasars with 
bolometric luminosities $L_{bol}$ to the luminosity budget should be 
augmented by roughly $1/(1-f^2(L_{bol}))$ to account for the obscured 
sources. Other recent calculations are based on the hard X-ray luminosity 
functions of AGN \citep{marc04, barg05} in an effort to include both 
obscured and unobscured AGN, but this procedure misses the contribution 
from Compton-thick sources. The most recent accretion efficiency estimates 
are based on combining optical and X-ray data to produce bolometric 
luminosity functions \citep{hopk07, shan07}. Nevertheless, the fraction of 
Compton-thick sources which would be missed by both X-ray surveys and type 
1 AGN surveys remains a major uncertainty in these methods. If 
$f_C(L_{bol})$ is the Compton-thick fraction of type 2 AGN, then the 
contribution of AGN with bolometric luminosities $L_{bol}$ should be 
increased by $1/(1-f_C(L_{bol})f(L_{bol}))$. Assuming $f_C\ga 0.5$ \citep{risa99} and $f\ga 0.5$ (this work) for quasars around and above the luminosity function break, we find that both types of methods (those based on the optically selected AGN and those based on X-ray selected AGN) may underestimate the AGN luminosity budget by 30\% or more. This means that the calculated accretion efficiency would be underestimated by the same amount.

\section{Summary and Conclusions}
\label{sec:conclusions}

In this paper, we present the largest sample of type 2 quasars to date. Table~\ref{tab:qso2} lists the 887 optically-selected sources with redshifts $z\lesssim 0.83$. These objects are selected from the spectroscopic database of the SDSS on the basis of their emission line properties. Candidate sources are required to have no broad (FWHM$>1100$ km/sec) components in their permitted emission lines. To distinguish type 2 AGN from star-forming galaxies, we use the standard line diagnostic diagrams involving [OIII]/H$\beta$, [NII]/H$\alpha$ and [SII]/H$\alpha$ line ratios (at redshifts $z<0.3$) or require other signs of a hidden AGN, such as the presence of the [NeV]3346,3426 emission lines (at redshifts $0.3<z<0.8$). We place a lower limit on the [OIII]5007 line luminosity of $10^{8.3}L_{\odot}$, ensuring that the bolometric luminosities of these objects are above the classical Seyfert/quasar separation of $10^{45}$ erg s$^{-1}$. 

For this sample, we calculate the \oiiil\ luminosity function using the 1/\vmax\ method (Fig.~\ref{fig:phitable_hao}). The selection function that we calculate for each object involves taking into account four different spectroscopic target selection algorithms, of which two (Low-z QSO and High-z QSO) are color-based, one (Galaxy) is based on optical morphology and one (Serendipity FIRST) is based on the radio properties. By comparing how well each algorithm performs at selecting type 2 quasars, we find that color-based algorithms are rather ineffective and select no more than 20\% of objects. Indeed, there is no well-defined region of the color-color space where type 2 quasars concentrate, even when only narrow redshift ranges are considered. The success of the SDSS in selecting a large number of type 2 quasars is due to the unprecedented size of the spectroscopic survey and the multitude of different target selection algorithms which allow for serendipitous objects. 

We extend the \oiiil\ AGN luminosity function to luminosities about 2 orders of magnitude higher than was previously done, up to $L_{\rmoiii}\simeq 10^{10}L_{\odot}$. This value corresponds to an intrinsic UV luminosity of $M_{2500}=-28$ mag and a bolometric luminosity of $4\times 10^{47}$ erg s$^{-1}$. 

We also derive the \oiiil\ luminosity function for a complete sample of 8,003 $z<0.83$ type 1 quasars taken from the SDSS quasar catalog and find it to be in excellent agreement with other measurements. We can then directly compare the luminosity functions of type 1 and type 2 quasars (Fig.~\ref{fig:phitable_qso12_union_zbin}) and constrain the type 2 quasar fraction as a function of luminosity (Fig.~\ref{fig:type2_frac}). We argue that the type 2 quasar luminosity function and the type 2 quasar fraction that we derive are robust lower limits. The main reasons are that (i) there may be a significant number of type 2 quasars that do not meet our spectroscopic selection criteria, and (ii) narrow lines are more extincted in type 2 quasars than they are in type 1 quasars, biasing the type 2/type 1 ratio at a given luminosity to lower values.

Objects with different \oiiil {} luminosities and different redshifts suffer from different selection biases. Our best data are at low redshifts and relatively low luminosities ($z<0.3$ and $L_{\rmoiii}<10^9L_{\odot}$) and at high redshifts and relatively high luminosities (\zc\ and $L_{\rmoiii}>10^{9.5}L_{\odot}$). In these regimes we find that type 2 quasars are more abundant than type 1 quasars, with the type 2/type 1 ratios of 1.5:1 and 1.2:1, correspondingly. 

\section*{Acknowledgments}
We would like to thank the referee, H. Netzer, for the helpful comments. NLZ is supported by the NASA Spitzer Space Telescope Fellowship Program through a contract issued by the Jet Propulsion Laboratory, California Institute of Technology under a contract with NASA. RR and MAS are supported by NSF grant AST-0707266. Support for this work was also provided by the NASA through Chandra X-ray Observatory Center, which is operated by the Smithsonian Astrophysical Observatory for and on behalf of the NASA under contract NAS8-03060. 

Funding for the SDSS and SDSS-II has been provided by the Alfred P. Sloan Foundation, the Participating Institutions, the National Science Foundation, the U.S. Department of Energy, the National Aeronautics and Space Administration, the Japanese Monbukagakusho, the Max Planck Society, and the Higher Education Funding Council for England. The SDSS is managed by the Astrophysical Research Consortium for the Participating Institutions. The Participating Institutions are the American Museum of Natural History, Astrophysical Institute Potsdam, University of Basel, Cambridge University, Case Western Reserve University, University of Chicago, Drexel University, Fermilab, the Institute for Advanced Study, the Japan Participation Group, Johns Hopkins University, the Joint Institute for Nuclear Astrophysics, the Kavli Institute for Particle Astrophysics and Cosmology, the Korean Scientist Group, the Chinese Academy of Sciences (LAMOST), Los Alamos National Laboratory, the Max-Planck-Institute for Astronomy (MPIA), the Max-Planck-Institute for Astrophysics (MPA), New Mexico State University, Ohio State University, University of Pittsburgh, University of Portsmouth, Princeton University, the United States Naval Observatory, and the University of Washington. 

\section*{Appendix}
In this appendix, we prove our assertion in \S\ref{subsec:vvmax_alpha} that the statistic $u$, defined by Eq.~\ref{eq_u}, follows a $\chi^2$ distribution with one degree of freedom.  
We have $N$ quasars with the same redshift and \oiiil {} luminosity, of which $n$ are detected in the radio,
so the measured detection rate is $f=n/N$. We would like to test the null hypothesis (NH) that the
underlying probability of detection (for this redshift and luminosity) is $p$ (denoted as $\pzl$ in the text). If it were indeed $p$, then
the probability of detecting $n$ objects out of $N$ would be binomial:
\begin{equation}
R(n)=C_N^n p^n (1-p)^{N-n}.
\end{equation}
For very large $N$ ($\sqrt{Np}\gg1$), this distribution is close to a Gaussian
\begin{equation}
r(n){\rm d}n=\frac{1}{\sqrt{2\pi\sigma^2}}e^{-\frac{(n-pN)^2}{2\sigma^2}}{\rm d}n,
\end{equation}
$\mbox{where }\sigma^2=Np(1-p)$. We now define a function of the observed values $n$ and $N$ and the underlying probability $p$:
\begin{equation}
u(n)=\frac{(n-pN)^2}{Np(1-p)}.
\end{equation}
The probability density of $u$, i.e., the probability to find this variable in the range between $u$ and
$u+{\rm d}u$ under the assumption of the NH is
\begin{equation}  \label{eq_chi2}
r_u(u){\rm d}u=\frac{2r(n)}{|{\rm d}u/{\rm d}n|}{\rm d}u=\frac{1}{\sqrt{2\pi u}}e^{-u/2}{\rm d}u
\end{equation}
for $u>0$, and $r_u(u)=0$ otherwise. In other words, the distribution of $u$ does not depend on $N$ or $p$ for large values of $N$. The
distribution function (\ref{eq_chi2}) is that of $\chi^2$ with one degree of freedom. Given a measurement
$n$ and $N$, if we want to test whether it is consistent with the underlying probability $p$, we calculate
$u$. If $u>6.5$, then the NH is ruled out with a 99\% probability.



\begin{deluxetable}{rrrrrrrrrrrrrrrrrrrrr}\rotate
\tabletypesize{\tiny}
\tablewidth{0pt}
\setlength{\tabcolsep}{0.03in}
\tablecaption{Catalog of 887 Optically-Selected Type 2 Quasars \label{tab:qso2}}
\tablehead{RA & Dec & redshift & $L_{\rm [OIII]}^{\rm (G)}$ & $L_{\rm [OIII]}^{\rm(NP)}$ & Plate & Fiber & MJD & Tcode & $S_{\rm 20 cm}$ & RMS & $u_{\rm PSF}$ & $g_{\rm PSF}$ & $r_{\rm PSF}$ & $i_{\rm PSF}$ & $z_{\rm PSF}$ & $\sigma_u$ & $\sigma_g$ & $\sigma_r$ & $\sigma_i$ & $\sigma_z$}
\startdata 
   0.746236 &   0.671701 & 0.6007 & 9.04 & 9.12 &  686 &  350 &  52519 & 00001 &    0.00 & 0.149 &21.67 &21.35 &20.54 &19.72 &19.83 &0.36 &0.05 &0.04 &0.03 &0.11 \\ 
   0.965958 &  $-1.028334$ & 0.2689 & 8.16 & 8.15 &  669 &  289 &  52559 & 00001 &    8.45 & 0.148 &21.93 &20.68 &19.26 &18.84 &18.31 &0.20 &0.03 &0.02 &0.02 &0.03 \\ 
   1.222618 &  $-0.843989$ & 0.6430 & 9.18 & 9.20 &  669 &  209 &  52559 & 00001 &    0.00 & 0.153 &22.37 &21.90 &21.03 &20.19 &19.97 &0.32 &0.11 &0.06 &0.04 &0.13 \\ 
   1.870121 &   1.101123 & 0.4663 & 8.33 & 8.33 &  669 &  457 &  52559 & 00001 &    0.00 & 0.084 &21.80 &21.80 &20.29 &19.66 &19.19 &0.21 &0.07 &0.03 &0.03 &0.07 \\ 
   2.799806 &   0.940632 & 0.4094 & 8.67 & 8.66 &  669 &  602 &  52559 & 00001 &    0.00 & 0.117 &21.89 &21.10 &19.84 &19.19 &18.99 &0.30 &0.06 &0.03 &0.02 &0.06 \\ 
   2.862281 &  15.891562 & 0.0999 & 8.16 & 8.17 &  752 &  380 &  52251 & 10000 &    0.00 &-1.000 &20.72 &19.28 &18.50 &17.88 &17.61 &0.09 &0.02 &0.02 &0.02 &0.03 \\ 
   3.026290 &  $-9.790457$ & 0.1668 & 8.54 & 8.55 &  652 &  399 &  52138 & 10000 &    0.00 & 0.153 &20.42 &18.95 &18.16 &17.81 &17.69 &0.07 &0.02 &0.02 &0.02 &0.03 \\ 
   5.070319 &  $-9.545760$ & 0.3600 & 8.60 & 8.62 & 1913 &  381 &  53321 & 00001 &    0.00 & 0.149 &20.05 &19.46 &18.74 &18.54 &18.27 &0.05 &0.02 &0.01 &0.02 &0.03 \\ 
   5.344129 &  $-0.254834$ & 0.5493 & 8.35 & 8.36 &  687 &   22 &  52518 & 00001 &    1.48 & 0.175 &21.35 &21.02 &20.24 &19.82 &19.61 &0.15 &0.04 &0.03 &0.03 &0.08 \\ 
   6.381066 & $-10.672835$ & 0.3035 & 8.64 & 8.65 &  653 &  149 &  52145 & 01000 &    1.16 & 0.135 &19.78 &19.59 &19.06 &18.94 &18.50 &0.05 &0.02 &0.02 &0.02 &0.04 
\enddata 
\tablecomments{(1) RA and Dec are in J2000.0 coordinates. \\
(2) $L_{\rm [OIII]}^{\rm (G)}$ and  $L_{\rm [OIII]}^{\rm (NP)}$ are listed in units of $\log(L/L_\odot)$ and refer to Gaussian and non-parametric measures of the \oiiil {} line luminosity, respectively (\S\ref{subsec:type2_o3}).\\
(3) The first four digits of the target code (col. 9) show whether the object was targeted with the Galaxy, Low-z QSO, High-z QSO, or Serendipity FIRST algorithm, in that order. The last digit indicates whether the object belongs to the Special Southern survey.\\
(4) $S_{\rm 20 cm}$ is the peak flux (in mJy/beam) of the nearest match within 2\arcsec {} from the FIRST survey \citep{beck95}, and is listed as zero, if there is no match. If the object is not in the FIRST survey area, the RMS flux value is listed as $-1$.\\
(5) Columns 12-16 and 17-21 list $ugriz$ PSF asinh magnitudes from the SDSS TARGET database, corrected for Galactic extinction \citep{schl98}, and their 1-sigma errors.\\
Table 1 is presented in its entirety in the electronic edition of the Astronomical Journal. A portion is presented here for guidance regarding its form and content.}
\end{deluxetable}

\begin{deluxetable}{lc}
\tabletypesize{\tiny}
\tablewidth{0pt}
\setlength{\tabcolsep}{0.03in}
\tablecaption{SDSS Spectroscopic Target Algorithms \label{tab:target}}
\tablehead{Target Algorithm & No. of Obj.}
\startdata
\multicolumn{2}{l}{\textbf{Main Survey}} \\
\hline
Main Galaxy & 366\\
Main Low-z QSO & 76\\
Main High-z QSO & 89 \\
Main Serendipity FIRST & 276\\
Main LRG & 64 \\
Main QSO FIRST & 14 \\
Main ROSAT & 28 \\
\hline
\textit{Combined} & 771 \\
\hline
\multicolumn{2}{l}{\textbf{Special Southern Survey}} \\
\hline
Southern Galaxy & 4\\
Southern LRG & 5\\
Southern Low-z QSO & 20\\
Southern High-z QSO & 6\\
Southern Serendipity FIRST & 9\\
Southern ROSAT & 1\\
Faint quasars & 31\\
Photo-z & 29\\
Faint LRG & 5\\
$u$-band galaxy & 2\\
\hline
\textit{Combined} & 116
\enddata
\tablecomments{SDSS spectroscopic target algorithms important for selecting type 2 quasars, and the number of objects targeted by each. Objects can be targeted by multiple algorithms, so these numbers do not add up to the total number of objects. The type 2 quasar luminosity function presented in this paper is derived from the 740 objects targeted by the top four Main target algorithms: Galaxy, Low-z QSO, High-z QSO and Serendipity FIRST.}
\end{deluxetable}

\begin{deluxetable}{lcc}
\tabletypesize{\small}
\tablewidth{0pt}
\tablecaption{Mean \vvmax {} values calculated from the type 1 and type 2 quasar samples. \label{tab:vobsvmax}}
\tablehead{ & No. & \mvvmax }
\startdata
\multicolumn{3}{c}{Type 1 Quasars}\\
\hline
$0.00<z\le 0.30$ & 1020 & 0.55\\
$0.30<z\le 0.50$ & 2802 & 0.56\\
$0.50<z<0.83$ & 4181 & 0.57\\
Combined & 8003 & 0.61 \\
\hline
\multicolumn{3}{c}{Type 2 Quasars}\\
\hline
$0.00<z\le 0.30$ & 420 & 0.46\\
$0.30<z\le 0.50$ & 175 & 0.61\\
$0.50<z<0.83$ & 145 & 0.54\\
\enddata
\tablecomments{Our type 1 quasar sample is complete and values greater than 0.5 indicate that there is positive redshift evolution within each redshift range. Our type 2 quasar sample is incomplete and the deviation of \mvvmax {} from 0.5 is due to a combination of redshift evolution and selection effects, which we do not attempt to disentangle.}
\end{deluxetable}

\end{document}